\documentclass[manuscript]{aastex6}
\usepackage{amsmath}
\usepackage{color}
\usepackage{comment}
\usepackage{url}

\usepackage{graphicx}

\begin{document}
\shorttitle{Statistical Studies of Solar White-Light Flares}
\shortauthors{Namekata et al.}


\title{Statistical Studies of Solar White-Light Flares and Comparisons with Superflares on Solar-type Stars}






\author{Kosuke Namekata\altaffilmark{1}}

\author{Takahito Sakaue\altaffilmark{1} }
\author{Kyoko Watanabe\altaffilmark{2} }
\author{Ayumi Asai\altaffilmark{3,4} }
\author{Hiroyuki Maehara\altaffilmark{5} }
\author{Yuta Notsu\altaffilmark{1} }
\author{Shota Notsu\altaffilmark{1} }
\author{Satoshi Honda\altaffilmark{6} }
\author{Takako T. Ishii\altaffilmark{3}}
\author{Kai Ikuta\altaffilmark{1}}
\author{Daisaku Nogami\altaffilmark{1} }
\and

\author{Kazunari Shibata\altaffilmark{3,4}}


\altaffiltext{1}{Department of Astronomy, Kyoto University, Kitashirakawa-Oiwake-cho, Sakyo-ku, Kyoto 606-8502, Japan; namekata@kusastro.kyoto-u.ac.jp}
\altaffiltext{2}{National Defense Academy of Japan, 1-10-20 Hashirimizu, Yokosuka, 239-8686, Japan}
\altaffiltext{3}{Kwasan and Hida Observatories, Kyoto University, Yamashina, Kyoto 607-8471, Japan}
\altaffiltext{4}{Unit of Synergetic Studies for Space, Kyoto University, Yamashina, Kyoto 607-8471, Japan}
\altaffiltext{5}{Okayama Astrophysical Observatory, National Astronomical Observatory of Japan, Asaguchi, Okayama, 719-0232, Japan}
\altaffiltext{6}{Nishi-Harima Astronomical Observatory, Center for Astronomy, University of Hyogo, Sayo-cho, Sayo-gun, Hyogo, 679-5313, Japan}

\begin{abstract}
Recently, many superflares on solar-type stars have been discovered as white-light flares (WLFs).
The statistical study found a correlation between their energies ($E$) and durations ($\tau$): $\tau \propto E^{0.39}$ \citep[][EP\&S, 67, 59]{2015EP&S...67...59M}, similar to those of solar hard/soft X-ray flares: $\tau \propto E^{0.2-0.33}$. 
This indicates a universal mechanism of energy release on solar and stellar flares, i.e., magnetic reconnection. 
We here carried out a statistical research on 50 solar WLFs observed with \textit{SDO}/HMI and examined the correlation between the energies and durations.
As a result, the $E$--$\tau$ relation on solar WLFs ($\tau \propto E^{0.38}$) is quite similar to that on stellar superflares ($\tau \propto E^{0.39}$). 
However, the durations of stellar superflares are one order of magnitude shorter than those expected from solar WLFs. 
We present the following two interpretations for the discrepancy.
(1) In solar flares, the cooling timescale of WLFs may be longer than the reconnection one, and the decay time of solar WLFs can be elongated by the cooling effect.
(2) The distribution can be understood by applying a scaling law ($\tau \propto E^{1/3}B^{-5/3}$) derived from the magnetic reconnection theory.
In this case, the observed superflares are expected to have 2-4 times stronger magnetic field strength than solar flares.
\end{abstract}


\section{Introduction} \label{sec:int}
Solar flares are abrupt brightenings on the solar surface. 
During flares, magnetic energy stored around sunspots is believed to be converted to kinetic and thermal energies through the magnetic reconnection in the corona \citep[e.g.,][]{1981sfmh.book.....P,2011LRSP....8....6S}.
In the standard scenario, the released energies are transported from the corona to the lower atmosphere by nonthermal high energy particles and  thermal conduction.
The energy injection causes chromospheric evaporations and chromospheric condensations, which are observed as bright coronal and chromospheric emission, respectively.

Flares in the visible continuum are particularly called white-light flares (hereinafter WLFs), firstly observed by \cite{1859MNRAS..20...13C} .
Solar WLFs are usually rare events compared to the $\rm H\alpha$ and soft X-ray flares because of the short durations \citep[typically a few minutes; ][]{1992PASJ...44L..77H,2006ApJ...641.1210X} and the low contrast \citep[typically 5-50\%, at most 300\%; ][]{1976SoPh...50..153L,2008ApJ...688L.119J}.
It is widely accepted that white-light (WL) emissions are well correlated with hard X-ray and radio emissions spatially \citep[e.g., ][]{2011ApJ...739...96K} and temporally \citep[e.g., ][]{2006SoPh..234...79H}.
These properties imply that high energy electrons are essential to the WL emissions.
However, while the high energy electrons can penetrate into the chromosphere, they hardly reach the photosphere where optical photospheric continua originate \citep{1970SoPh...15..176N}.
For this reason, some portions of WL emissions are considered to be radiated through hydrogen recombination continuum (Paschen) at the directly heated and ionized upper chromosphere \citep{1989SoPh..124..303M,1976sofl.book.....S,2017ApJ...847...48H}.
On the other hand, the absence of strong chromospheric Balmer-continuum emission in some WLFs \citep[so-called ``type-II'' WLFs,][]{1986A&A...159...33M} is considered to suggest another WL emission source \citep{1999ApJ...512..454D}, which is $\rm H^{-}$ continuum from the heated photosphere \citep[also see][for review]{1989SoPh..124..303M,1976sofl.book.....S,2017ApJ...847...48H}.
The energy transportation to such a lower atmosphere is still debated. A strong downward irradiation by XUV or hydrogen Balmer continuum is proposed to heat the lower atmosphere ( the so-called ``back-warming'').
In addition to this, other energy injections have been also proposed, such as \textit{Alfv\'en} waves \citep{2008ApJ...675.1645F} or high energy protons \citep{1989SoPh..121..261N}.

In any case, the lack of information of the emission heights and spectra makes it difficult to identify the emission mechanism.
Although many authors have reported the emission heights compared to hard X-ray \citep[e.g.,][]{2012ApJ...753L..26M} and chromospheric line emission \citep{2013ApJ...776..123W}, there is no agreement due to its difficulty in observation.
The broad-band spectroscopic observations of solar WLFs are quite sparse \citep[e.g.,][]{1990ApJS...74..609M}, but enable us to fit the continuum with blackbody radiation. 
Some studies reported the emission temperature of 5,000--6,000 K \citep[e.g., ][]{2013ApJ...776..123W,2014ApJ...783...98K,2016ApJ...816...88K}, and the others 9,000 K \citep{2011A&A...530A..84K}.
The latter is similar to stellar observations (9,000-10,000 K) of flares on M-type stars \citep[e.g., ][]{1992ApJS...78..565H}.
The blackbody fitting would enable us to roughly estimate the radiation energies in the optical continuum, but it does not consider the Balmer recombination continuum whose large enhancement is proposed by some observations \citep[e.g.,][]{2014ApJ...794L..23H} and simulations \citep[e.g., ][]{2017ApJ...837..125K}.

In contrast to solar observations, the recent space-based optical telescopes found many stellar WLFs.
Interestingly, they also discovered many large WLFs, called ``superflares'', on solar-type stars (G-type main sequence stars) whose energies ($10^{33-36}$ erg) are 10--10,000 times larger than those of the maximum solar flares ($\sim10^{32}$ erg) \citep{2012Natur.485..478M,2013ApJS..209....5S}.
Various statistical studies show common properties between the solar and stellar flares as well as those between sunspots and star spots.
The occurrence frequency of flares and spots are universally expressed with the same power-law relation among the Sun and superflare stars \citep{2013ApJS..209....5S,2017PASJ...69...41M}, which implies a common energy storage mechanism.
The released energies through solar flares and superflares are found to be comparable with the magnetic energies stored around the spots \citep{2013ApJ...771..127N}.
This indicates that superflares are also phenomena where magnetic energies are released.

\cite{2015EP&S...67...59M} reported that there is a correlation between the energies radiated in WL ($E$) and durations ($\tau$) of superflares: $\tau \propto E^{0.39}$.
They found that the power-law relation is surprisingly consistent with those of solar flares observed with hard/soft X-rays: $\tau \propto E^{0.2-0.33}$ \citep{2002A&A...382.1070V,2008ApJ...677.1385C}.
A similar relation is also found between stored magnetic energies ($E_{\rm mag}$) and soft X-ray flare decay time ($\tau_{\rm decay}$): $\tau_{\rm decay} \propto E_{\rm mag}^{0.41}$ \citep{2017ApJ...834...56T}.
These similarities on the $E$--$\tau$ relation indicate a common mechanism of energy release among solar flares and supreflares.
\cite{2015EP&S...67...59M} moreover suggested that the observed power law relations can be explained by the magnetically driven energy release mechanism (magnetic reconnection) as follows.
Since flares are the mechanism that releases stored magnetic energies ($E_{\rm mag}$), flare energy ($E$) is expressed as a function of magnetic field strength ($B$) and length scale ($L$) of flares:
\begin{eqnarray}\label{eq:1}
E \sim fE_{\rm mag}\sim fB^2L^3,
\end{eqnarray}
where $f$ is a fraction of energy released by a flare.
On the other hand, the duration of flares ($\tau$) is thought to be comparable to the reconnection time scale ($\tau_{\rm rec}$):
\begin{eqnarray}\label{eq:2}
\tau \sim \tau_{\rm rec} \sim \tau_{A}/M_A \propto L/v_{A}/M_A,
\end{eqnarray}
where $\tau_A=L/v_{A}$ is the \textit{Alfv\'en} time, $v_A$ is the \textit{Alfv\'en} velocity, and $M_A$ is the dimensionless reconnection rate which takes the value of 0.1-0.01 in the case of the Petschek-type fast reconnection \citep{2011LRSP....8....6S}.
Assuming that stellar properties ($B$ and $v_A$) are not so different among the same spectral-type stars (solar-type stars), the values of both $E$ and $\tau$ are determined by the length scale ($L$).
On the basis of this assumption, the relation between $E$ and $\tau$ can be derived by deleting $L$ from Equation \ref{eq:1} and \ref{eq:2}:
\begin{eqnarray}\label{eq:3}
\tau \propto E^{1/3}.
\end{eqnarray}
This similarity between the theory and solar and stellar flare observations indirectly indicates that solar and stellar flares can be explained by the same mechanism of magnetic reconnection.

To confirm observationally this suggestion, a key is to investigate whether both solar and stellar flares are on a $E$--$\tau$ relation in a common wavelength range (X-ray or WL). 
In this study, we carried out a statistical research on 50 solar WLFs and compared the $E$--$\tau$ relations in the WL wavelength range, aiming to confirm the above expectation that solar and stellar flares can be universally explained by the same theoretical relation (Equation \ref{eq:3}).
We introduce analysis methods in Section \ref{sec:ana}, show the results in Section \ref{sec:res} and discuss the obtained results in Section \ref{sec:dis}

\section{Analysis} \label{sec:ana}
We carried out statistical analyses of temporal variations of 50 solar WLFs observed by \textit{Solar Dynamics Observatory} (\textit{SDO})/Helioseismic and Magnetic Imager \citep[HMI;][]{2012SoPh..275..207S} in the continuum channel with 45 sec cadence.
It is necessary to carefully subtract the background trend since the emission of solar WLFs is difficult to detect due to much lower contrast to the photosphere. 
On the basis of the previous studies \citep[e.g.,][]{2016ApJ...816....6K,2003A&A...409.1107M}, we identified the WL emissions inside the region with strong HXR emissions observed by \textit{RHESSI} \citep[\textit{Reuven Ramaty High Energy Solar Spectroscopic Imager;}][]{2002SoPh..210....3L}.
The radiated energies and decay times were calculated from the extracted light curves of WLFs. 
In the following subsections, we introduce these analysis methods.

\subsection{Selection} \label{subsec:21}
Our WLF catalogue contains M and X class solar flares which occurred from 2011 to 2015 and were observed by both \textit{SDO}/HMI and \textit{RHESSI}. 
The 43 flares in our catalogue which occurred from 2011 to 2014 were taken from \cite{2016ApJ...816....6K}, we enlarged the sample by adding 10 flares which occurred in 2015. 
The newly added flares satisfy the following three conditions.
\begin{itemize}
\item The \textit{GOES} X-ray class is above M2.
\item The overall light curve of a flare was observed by \textit{RHESSI}.
\item The WL emissions can be clearly recognized with the pre-flare-subtracted images.
\end{itemize}
The reason why we added only above M2 class flares is that the increase in numbers of weak flares would not improve statistics due to the difficulty in measurements of the emission.
After the imaging process of HXR with the \textit{RHESSI} data, we finally selected 50 solar flares (Table 1) whose HXR emissions are well spatially correlated with the WLF emissions.

\subsection{Time Evolution of WLF} \label{subsec:22}
As mentioned above, it is necessary to appropriately decide the area where we measure the enhancement of WLF to avoid the disturbance from granule or p-mode ocsillations. 
In this study, we extracted WL emissions inside the area identified by the HXR emissions observed with \textit{RHESSI}. 
In the imaging process of the \textit{RHESSI} data, we used the CLEAN algorithm \citep{2002SoPh..210...61H}. 
Detectors were carefully chosen on the basis of the \textit{RHESSI} detector spectrum (see, Table 2). 
The energy band was determined to be 30-80 keV because the HXR around 50 keV is well correlated with WLF emissions \citep[e.g.,][]{2016ApJ...816....6K}.
The integration time ranges were set from the beginning to the end of HXR flares.
As for the events whose loop-top sources have too strong HXR emissions, the initial impulsive phases were excluded to make clear maps of footpoint sources.

Comparing the images of the HXR flares and WLFs, we decided to make the light curves of WLFs by summing up the brightness variations of the HMI images inside the 30$\%$ contour of HXR emissions. 
This contour level is large enough to cover the enhancements of WLF emissions.
In order to make the light curves with high precision, the contour levels were adjusted to 10$\%$, 20$\%$, 50$\%$, or 70$\%$ for some events whose HXR emissions are too large or too small (Table 2). 
We determined the WLF's start and end time on the basis of the obtained light curves ($L_{\rm obs}$) and the pre-flare-subtracted WL movies (Table 2).
We subtracted the global trend as follows. 
First, we replaced the light curves during the flares with linear interpolation of the quiescence ($L_{\rm bg}$). 
Second, we obtained global background trends ($L_{\rm bg-trend}$) by performing a smoothing process (across 5 points of data) for the obtained background light curves ($\rm L_{bg}$).
Lastly, we subtracted the global background trends ($L_{\rm bg-trend}$) from the original light curves ($L_{\rm obs}$) and obtained the light curves of WLFs ($L_{\rm WL} = L_{\rm obs}-L_{\rm bg-trend}$). 
This analysis could hide meaningful trends such as increase in photospheric activity or gradual cooling components.
Considering these effects, we estimated the expected error bars in the Section \ref{subsec:32}.

Figure \ref{fig:1} shows the examples of the time evolutions of WLFs. 
(a) is one of the most powerful WLFs in our catalogue which occurred on 23th October 2012 (a X1.8 class flare). 
In the panel, light curves observed by \textit{GOES}, \textit{RHESSI} and \textit{SDO}/HMI are plotted in the left, and the pre-flare-subtracted images of the evolution of the WLF in the right. 
In the images, the time are indicated with arrows in the \textit{SDO}/HMI light curve, and the black components show the WL emissions over 1$\sigma$ levels. 
(b) and (c) show one of the long and short WLFs in our catalogue, respectively.
(d) is a disk-center event, and the WL kernel can be seen in the pre-flare subtracted images in the HXR contour.
As one can see, the main enhancement can be detected inside the \textit{RHESSI} contours, and the decay phases are usually a few minutes, as many previous studies also reported \citep[e.g.,][]{2016ApJ...816....6K,1992PASJ...44L..77H,2006ApJ...641.1210X,2003A&A...409.1107M}

\begin{figure}[p]
\begin{center}
\includegraphics[scale=0.5]{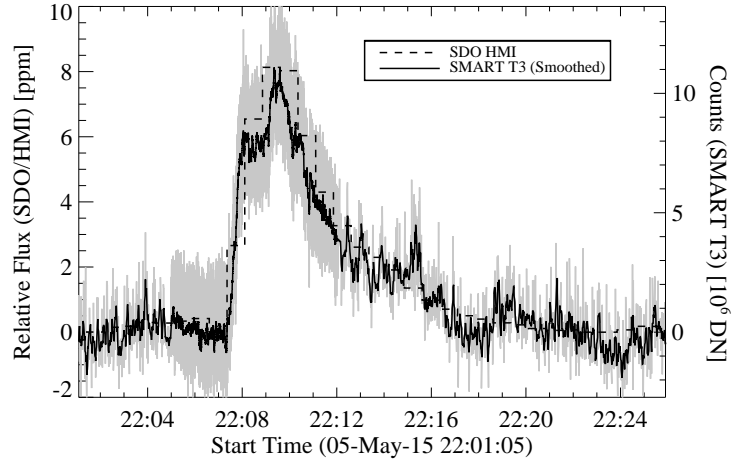}
\caption{Comparisons of the light curves of a solar white-light flare on 5th May 2015 observed by HMI continuum (6173 {\AA}; a dashed line) and SMART at Hida Observatory (continuum at 6470 {\AA}; a solid line). 
As for the SMART data, the gray solid line is the observed data, and the black one is the smoothed data.
SMART usually observes partial images of the Sun with 1 second cadence, but with 0.04 seconds cadence from 22:04:30 to 22:12:00 in the panel. 
Therefore, we obtained the black solid line by smoothing the gray line across 5 data points for the 1 second cadence data, and across 50 data points for 0.04 second cadence data.
Note that the vertical axes of the SMART and HMI data are fitted with the each peak value.
Although the 45 seconds cadence of HMI is worse than the 0.04 seconds (and 1 second) cadence of SMART, the light curve observed by HMI matches well that observed by SMART. The flare energy and duration (2.5$\times 10^{30}$ erg and 2.9 min) calculated by HMI data were comparable to those (2.3$\times 10^{30}$ erg and 2.5 min) calculated by SMART data. This indicates that the cadence of HMI is enough to follow the evolution of solar white-light flares.}
\label{fig:2}
\end{center}
\end{figure}

\begin{figure}[htbp]
\begin{center}
\includegraphics[scale=0.3]{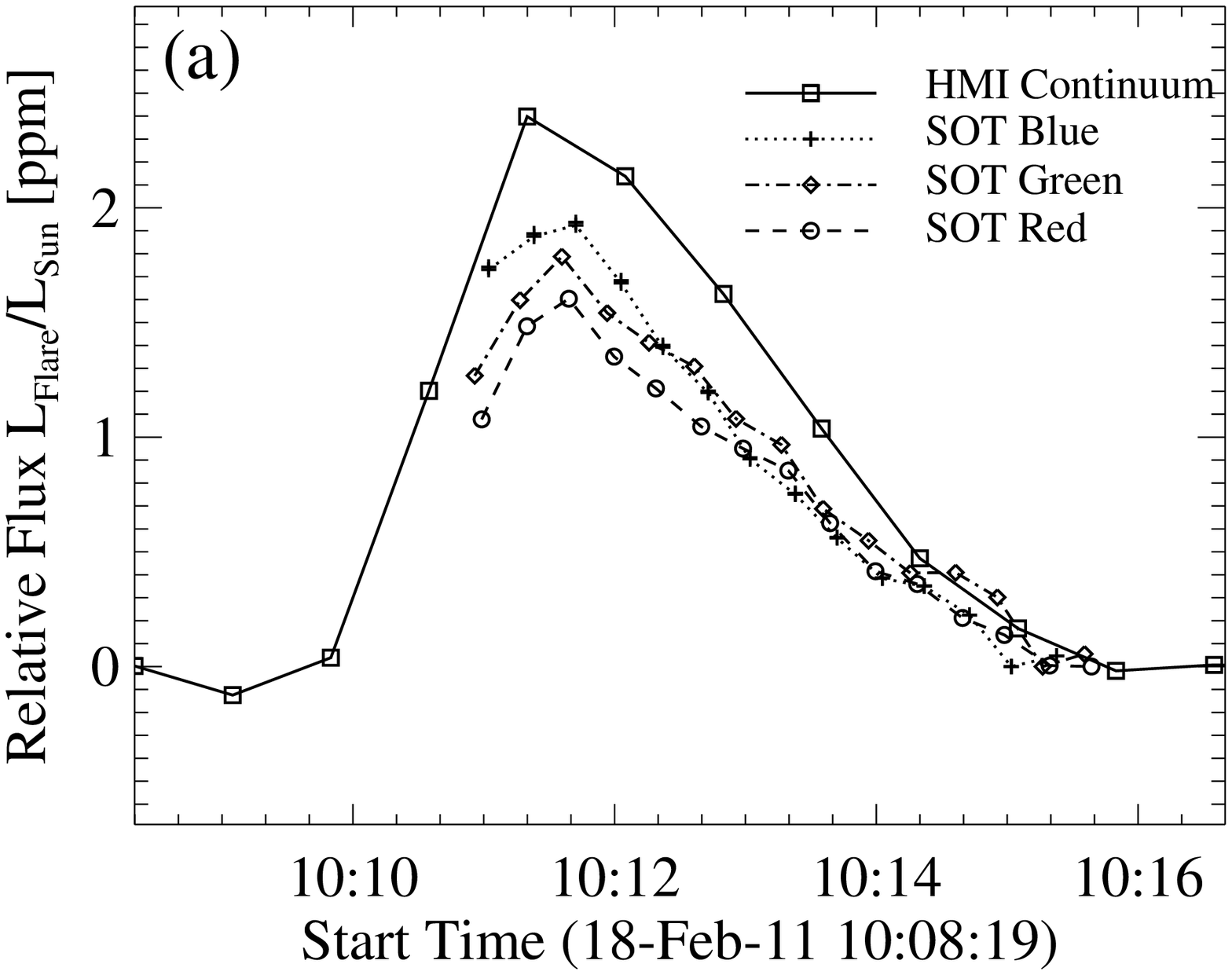}
\includegraphics[scale=0.3]{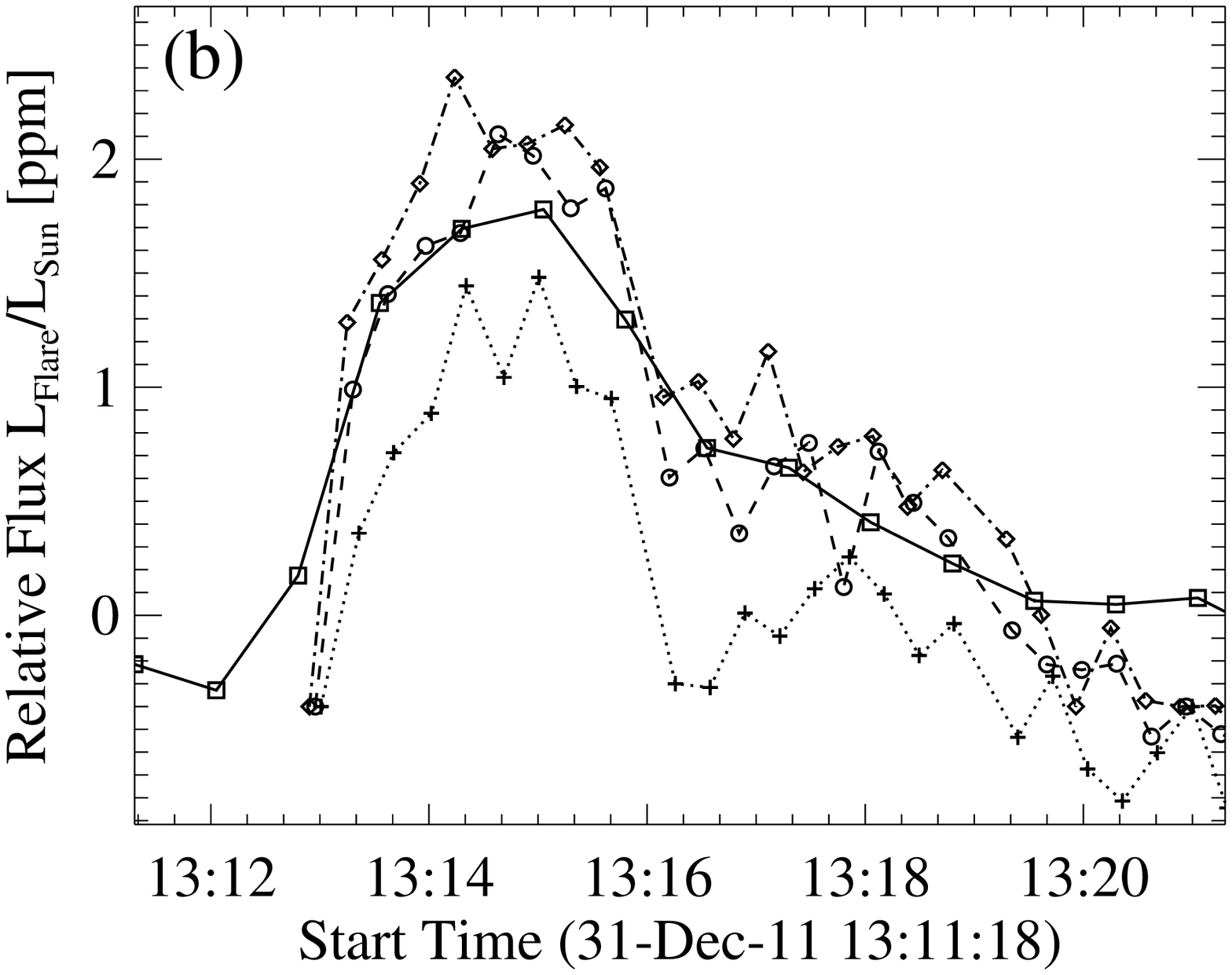}
\includegraphics[scale=0.3]{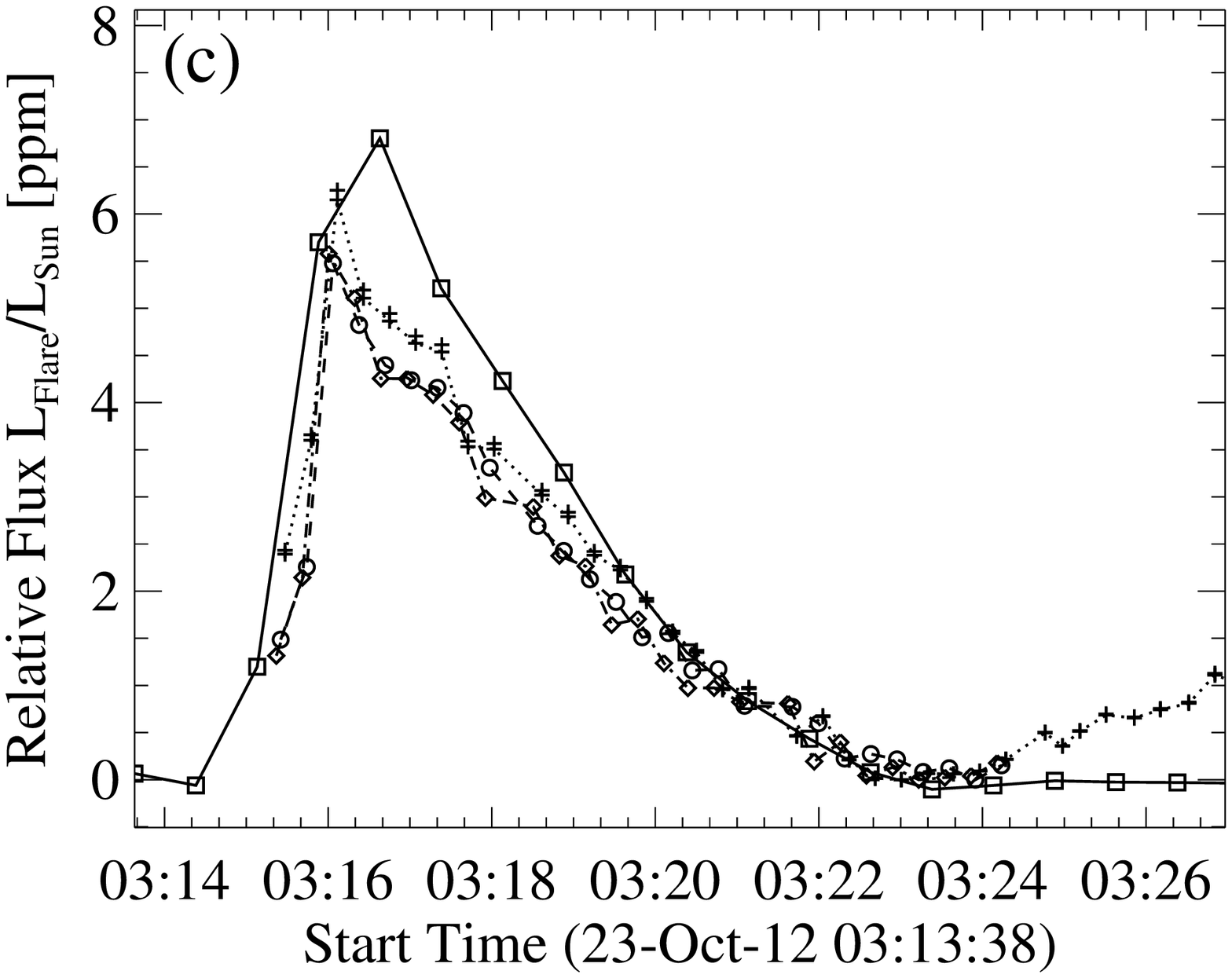}
\includegraphics[scale=0.3]{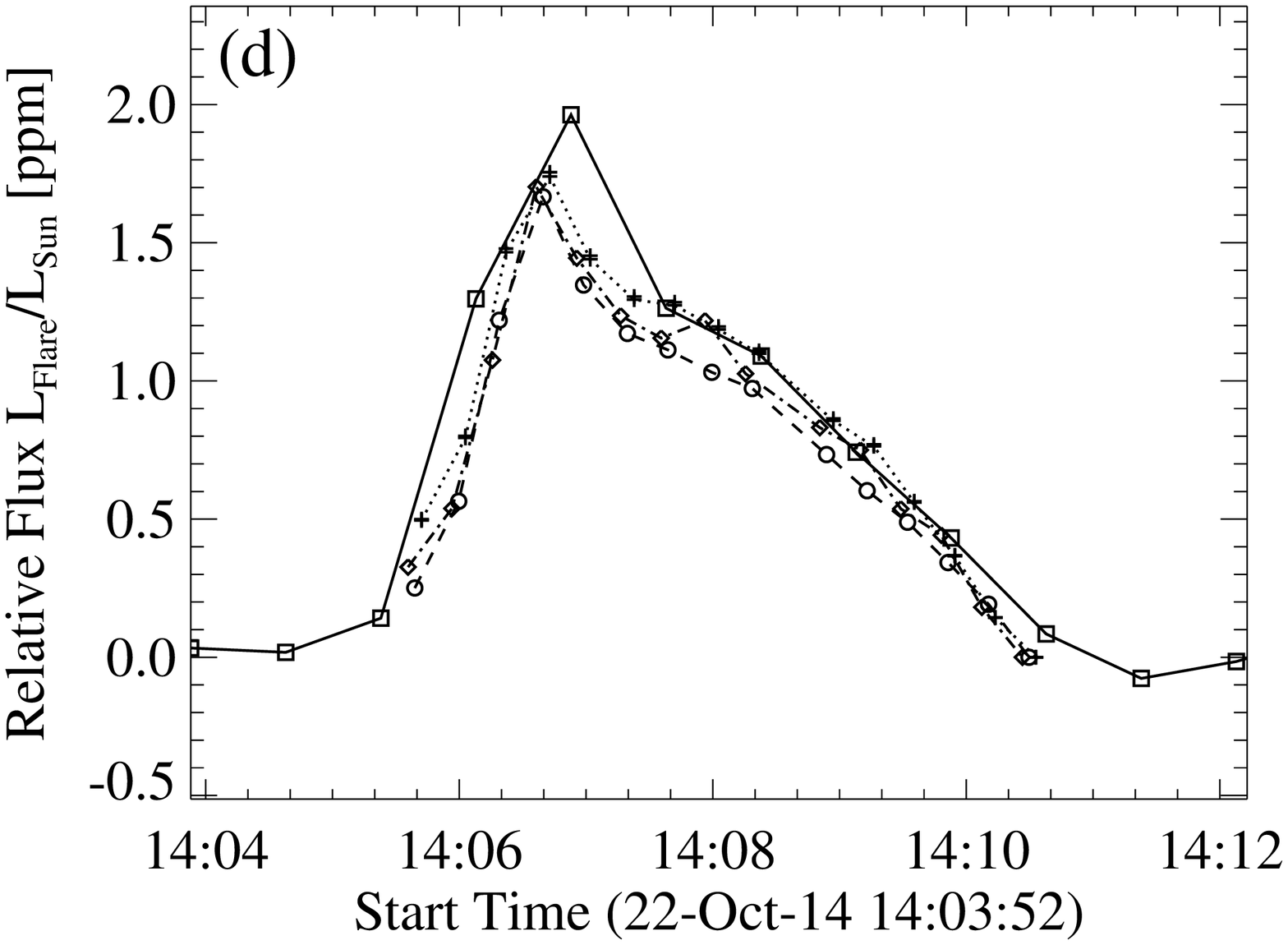}
\end{center}
\caption{Comparisons of the light curves of solar white-light flares observed with HMI continuum (open squares; solid lines) and with \textit{Hinode} SOT red (open circles; dashed lines), green (open diamonds; dash-dotted lines) and blue continuum (crosses; dotted lines). (a) is a flare on 18th Feb 2011, (b) a flare on 31th Dec 2011, (c) a flare on 23th Oct 2012, and (d) a flare on 22th Oct 2014.}
\label{fig:3}
\end{figure}

To present whether the cadence of \textit{SDO}/HMI is short enough to resolve the evolution of WLFs, 
we compared the obtained light curves with those observed by the \textit{Solar Magnetic Activity Research Telescope} \citep[\textit{SMART};][]{2013PASJ...65...39I} at Hida Observatory of Kyoto University for one event and \textit{Hinode}/Solar Optical Telescope \citep[SOT;][]{2008SoPh..249..167T} for four events.
In our catalogue, a flare on 5th May 2015 was also observed by \textit{SMART}.
The \textit{SMART} observes partial images of the Sun with a continuum filtergram (center 6470 {\AA}, width 10 {\AA}), and the time cadence is $\sim$0.04 sec.
We made a light curve by the same way as the above analysis and compared the time evolution with that obtained from HMI in Figure \ref{fig:2}.
These light curves, as in Figure \ref{fig:2}, look consistent with each other, which indicates that the cadence of HMI is enough to follow the evolution of solar WLFs.
There are also 4 events whose time profiles are clearly observed by \textit{Hinode}/SOT in the continuum channels (a red channel at 6684.0 {\AA}, and a green at 5550.5 {\AA}, a blue at 4504.5 {\AA}) among our catalogue (see, Table 3). 
Figure \ref{fig:3} shows that the obtained light curves well match each other.

\subsection{Calculation of Energy and Duration} \label{subsec:23}
\begin{figure}[htbp]
\begin{center}
\includegraphics[scale=0.28]{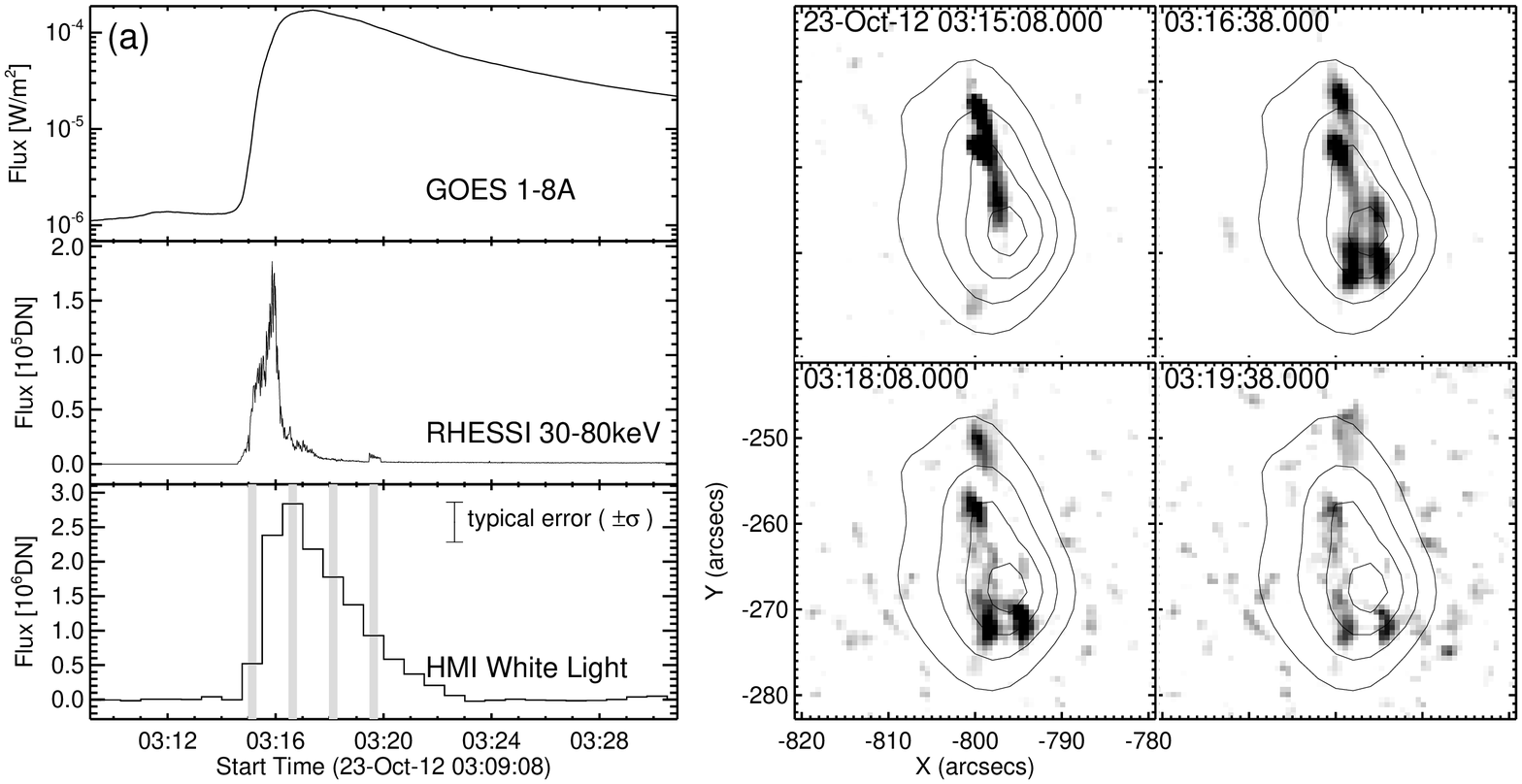}
\includegraphics[scale=0.28]{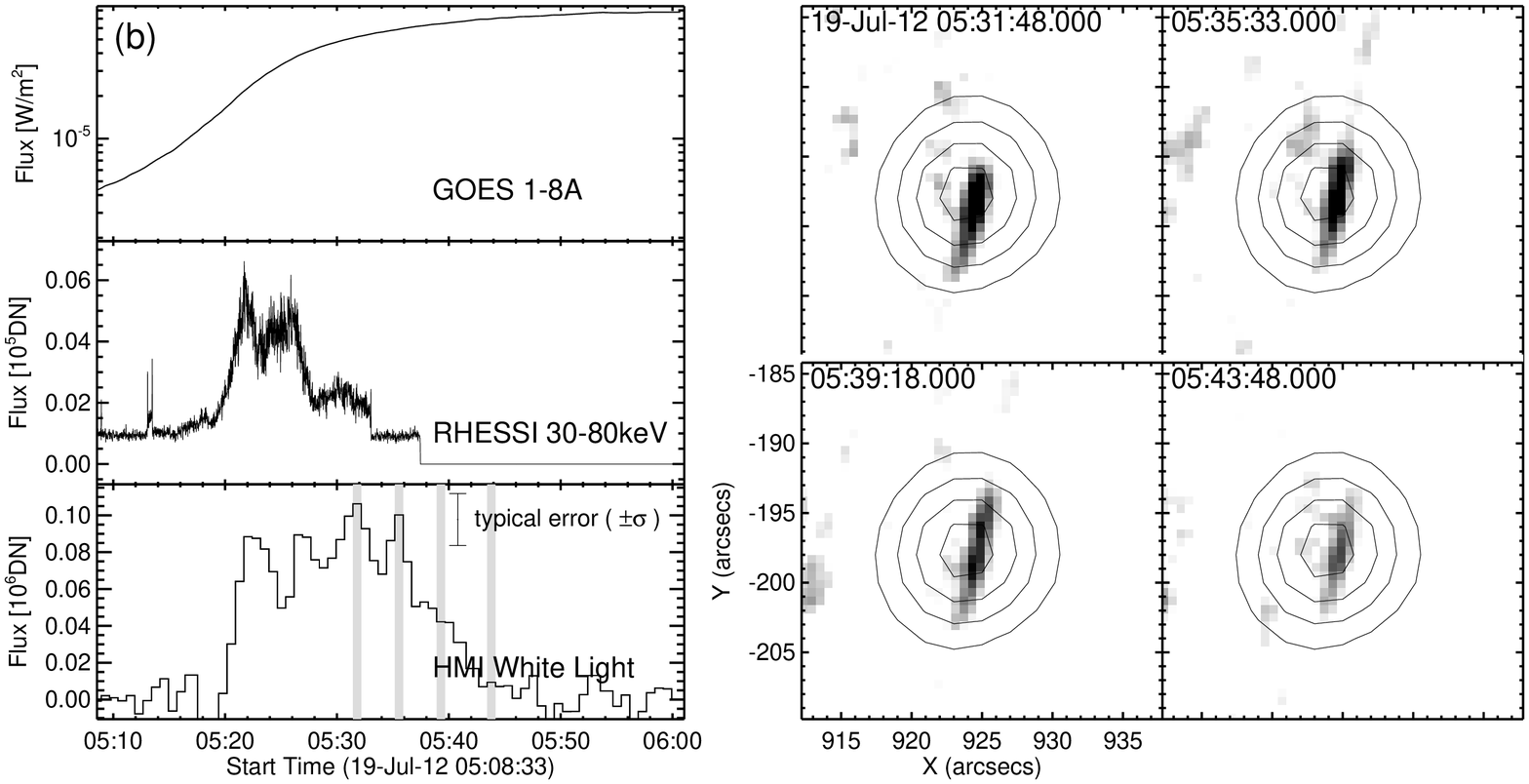}
\includegraphics[scale=0.28]{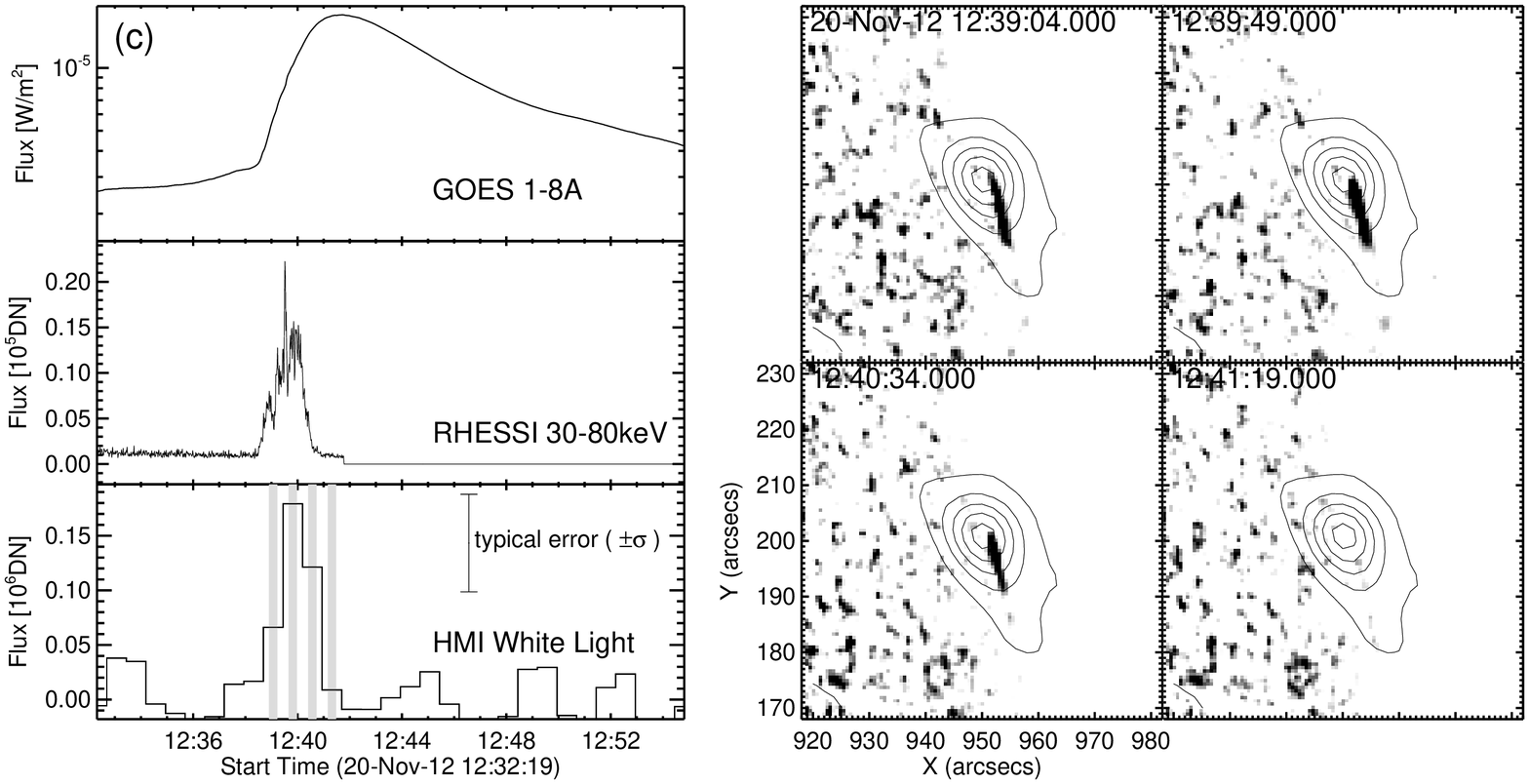}
\includegraphics[scale=0.28]{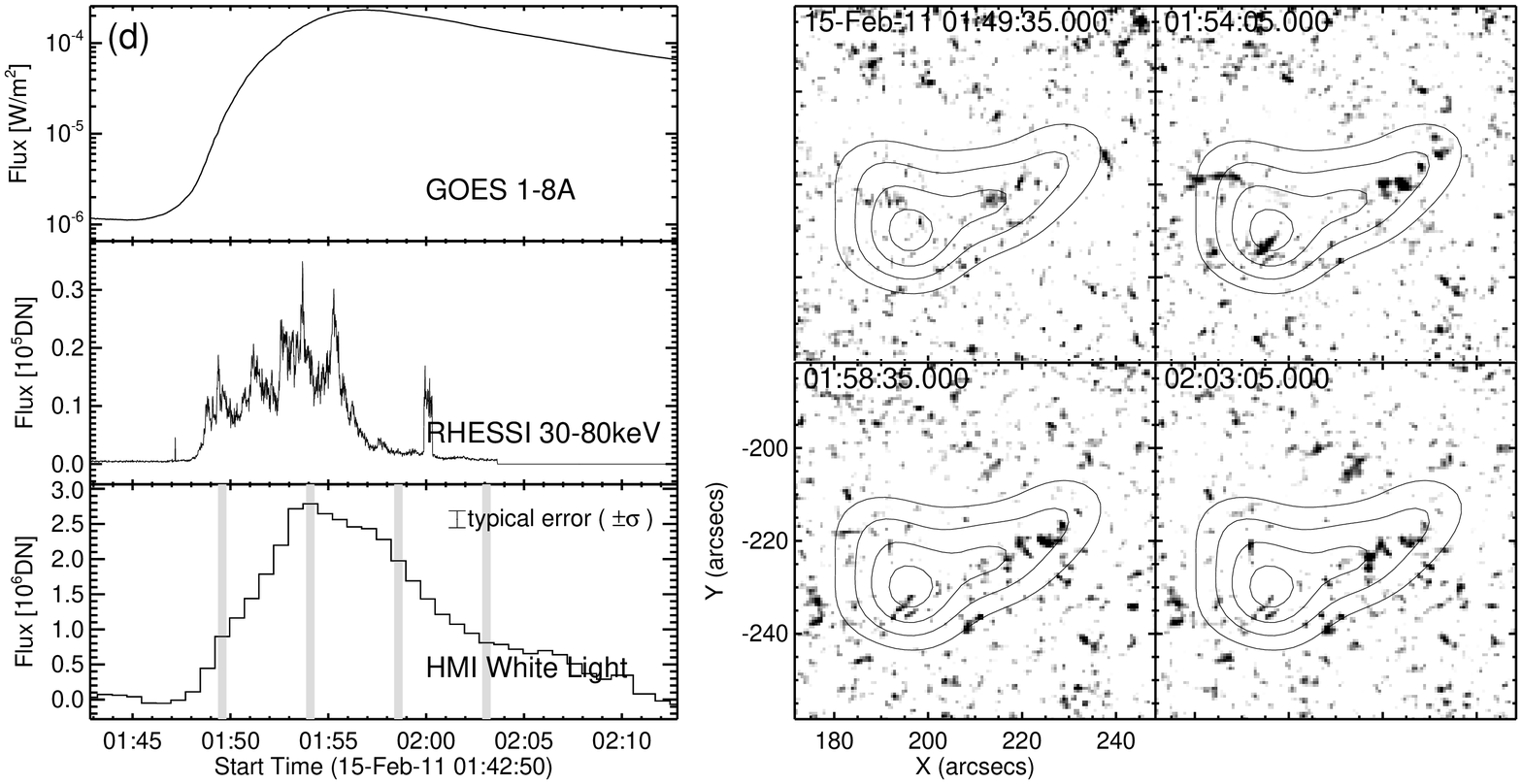}
\end{center}
\caption{(a) The left panels are the light curves of a solar flare which occurred on 23th October 2012 observed by \textit{GOES} (1-8 {\AA}; upper), \textit{RHESSI} (30-80keV; middle) and HMI (white light; lower). A typical error of WL emission is calculated on the basis of brightness variation outside the HXR contour of the HMI images. The right panels show the evolution of the pre-flare-subtracted images observed by HMI continuum and each time corresponds to that marked with gray line in the left lower panel. The black components show white-light emissions above 1$\sigma$ of each image and the black lines show the \textit{RHESSI} contours of 30$\%$, 50$\%$, 70$\%$ and 90$\%$ of the maximum emission in 30-80keV. (b) A long-duration solar flare on 19th July 2012, but the black lines show the \textit{RHESSI} contours of 10$\%$, 30$\%$ 50$\%$, 70$\%$ and 90$\%$. (c) A short-duration solar flare on 20th November 2012. (d) A disk-center solar flare on 15th Feb 2011.}
\label{fig:1}
\end{figure}

The main aim is to compare the energy ($E$) and duration ($\tau$) of solar and stellar WLFs.
We then calculated the energy and duration by the same way as \cite{2013ApJS..209....5S} and \cite{2015EP&S...67...59M}. 
There are two different things between solar and stellar observations, time cadences and pass bands. 
\textit{SDO}/HMI observes the overall Sun with a 45 sec cadence and a narrow-band filtergram around a 6173.3 {\AA} FeI line, while \textit{Kepler} carried out 1 min cadence observations with 4000-9000 {\AA} broad-band filters.
We calculated the energy of solar WLF assuming it is radiated by $T_{\rm flare}=$10,000 K blackbody \citep[see][for this assumption]{2015SoPh..290.3663K}:
\begin{eqnarray}
E &=& \sigma_{\rm SB\it}T_{\rm flare\it}^4\int A_{\rm flare\it}(t)dt \label{eq:ene}\\
A_{\rm flare\it}(t) &=& \frac{L_{\rm flare\it}}{L_{\rm sun\it}}\pi R^2 \frac{\int R_{\lambda}B_{\lambda}(5800\rm K\it)d\lambda}{\int R_{\lambda}B_{\lambda}(T_{\rm flare\it})d\lambda} \label{eq:are},
\end{eqnarray}
where $\sigma_{\rm SB\it}$ is \textit{Stefan-Boltzmann} constant, $L_{\rm flare\it}/L_{\rm sun\it}$ is the flare luminosity to the overall solar luminosity, $R$ is the solar radius, $R_{\lambda}$ is a response function of \textit{SDO}/HMI and $B_{\lambda}(T)$ is the \textit{Planck} function at a given wavelength $\lambda$. 
Since it is not confirmed how strongly the WL emission of flare is affected by the limb darkening effect \citep[e.g.,][]{2013ApJ...776..123W}, we obtained WL fluxes and energies with and without correcting for the limb darkening of a plane parallel atmosphere at 6000 {\AA}.
On the other hand, the durations of flares are also calculated as e-folding decay time of light curves as in \cite{2015EP&S...67...59M}. 
If we use the 45 sec cadence data, the decay time may be overestimated especially in the case of the flares with very short duration such as in the lower panels of Figure \ref{fig:1}.
We therefore calculated the decay time by using the light curves corrected for a linear interpolation of 1 sec cadence.

\section{Result} \label{sec:res}
\subsection{Relation between WLF Flux, Energy and SXR Flux} \label{subsec:31}
\begin{figure}[p]
\begin{center}
\plottwo{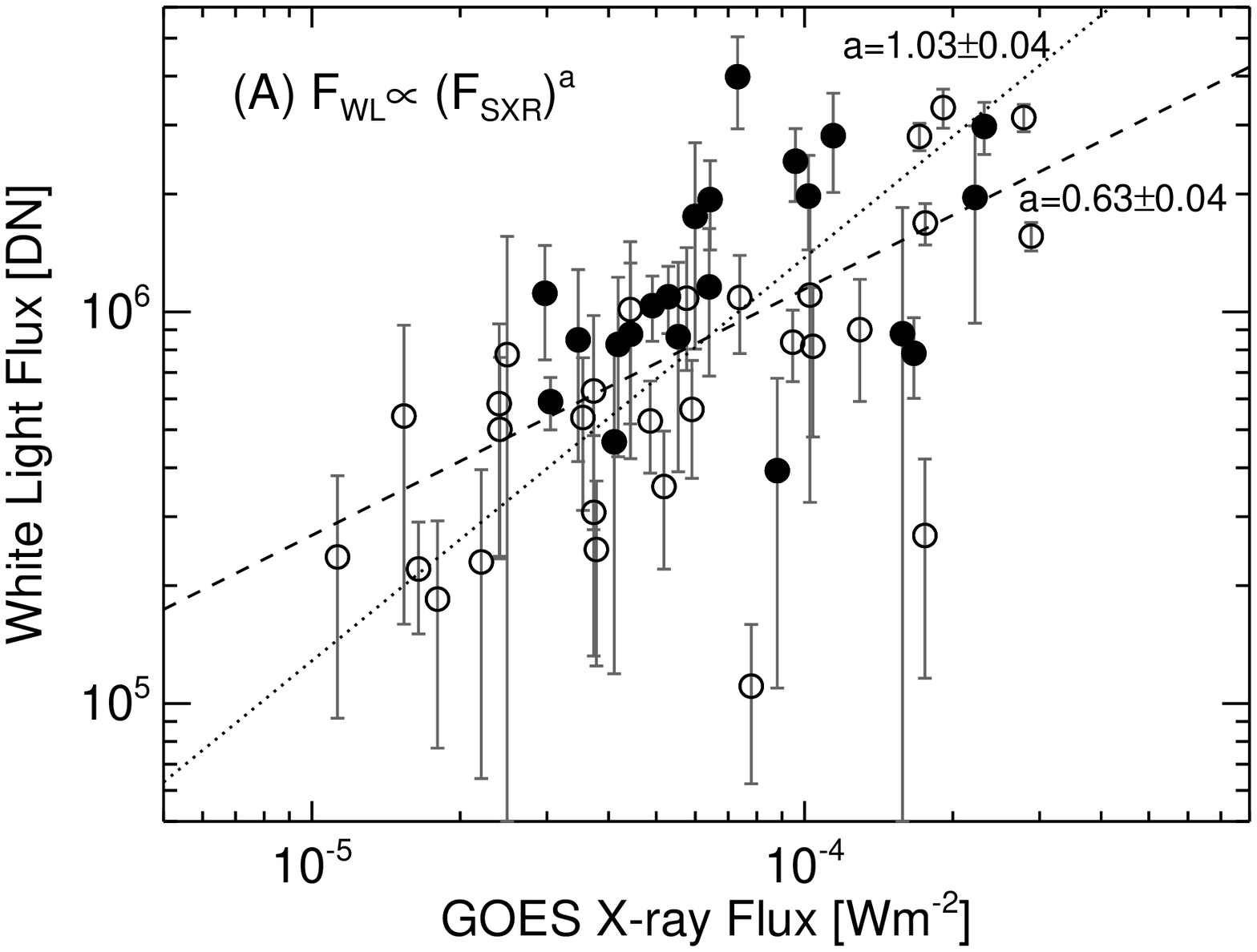}{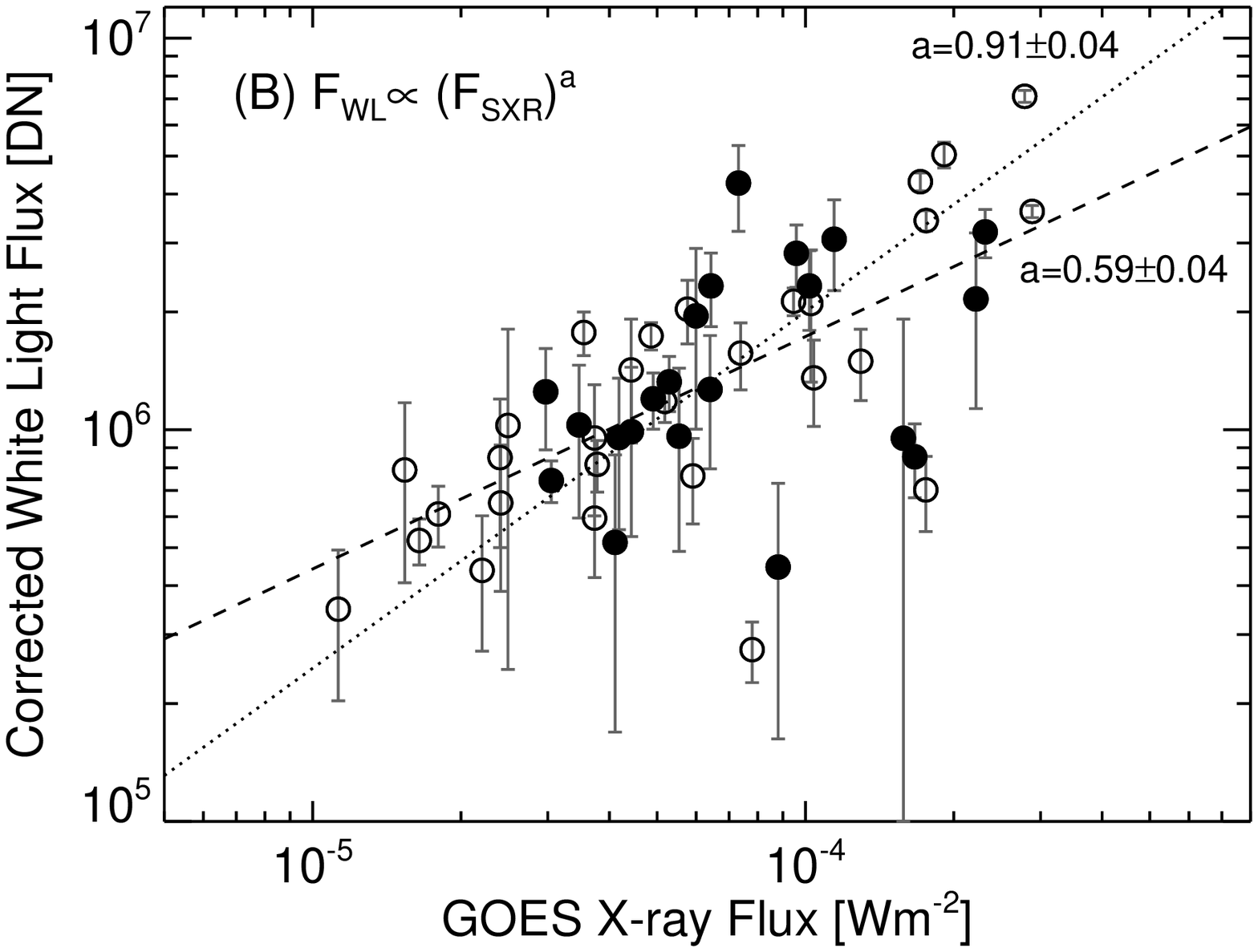}
\caption{(A) The left panel shows the comparisons of the GOES soft X-ray fluxes at the flare peak time and the HMI white-light fluxes. (B) The right panel shows the same as the left panel, but the white-light emissions are corrected assuming the limb-darkening of the plane-parallel atmosphere. In each panel, the filled symbols are flares on the disk center whose distances from the solar center are less than 700 arcsec, and open ones are flares on the limb. Dashed and dotted lines are fitted lines with linear regression method (LR) and a linear regression bisector method \citep[LRB; ][]{1990ApJ...364..104I}, respectively. The error bars shows 3$\sigma$ components of the light curve in quiescence.
The power-law indexes of the left panel are $a=0.63\pm 0.04$ (LR) and $a=1.03\pm 0.04$ (LRB), and those of the right panel are $a=0.59\pm 0.04$ (LR) and $a=0.91\pm 0.04$ (LRB).}
\label{fig:4}
\end{center}
\end{figure}

First, we show statistical properties concerning the WL fluxes, energies, and \textit{GOES} SXR fluxes.
In our catalogue, more than 40 flares were also analyzed by \cite{2016ApJ...816....6K}, and the values of WL fluxes obtained in this study are quite consistent with those obtained by \cite{2016ApJ...816....6K}.
The left panel in Figure \ref{fig:4} shows comparisons between WL fluxes ($F_{\rm WL}$) and \textit{GOES} soft X-ray fluxes ($F_{\rm SXR}$) at the flare peak, and the right panel shows the same but the WL fluxes are limb-darkening corrected fluxes. 
We fitted the relation between $F_{\rm WL}$ and $F_{\rm SXR}$ in the form of $F_{\rm WL}\propto (F_{\rm SXR})^{a}$ with the following two methods for comparison, a linear regression method (LR) and a linear regression bisector method \citep[LRB;][]{1990ApJ...364..104I}.
By the linear regression method, the power-law indexes of the left panel are $a=0.63\pm 0.04$, and those of the right panel are $a=0.59\pm 0.04$.

Likewise, the left panel in the Figure \ref{fig:5} shows comparisons between WL energies ($E_{\rm WL}$) and \textit{GOES} soft X-ray fluxes, and the right shows the same but the WL energies are the limb-darkening corrected ones. 
The power-law relations ($E_{\rm WL}\propto (F_{\rm SXR})^{b}$) are obtained with the indexes $b=0.87\pm 0.04$, and those of the right panel with $b=0.84\pm 0.04$.

\subsection{Relation between WLF Energy and Duration} \label{subsec:32}
Figure \ref{fig:6} shows comparisons between the radiated energies ($E$) and durations ($\tau$) of solar WLFs. 
The errors of flare durations are calculated by assuming that the pre-flare continuum levels fluctuated by 1$\sigma$.
There can be seen a positive correlation between flare energy and its duration and we can get the relation of $\tau \propto E^{0.38\pm 0.06}$ by fitting the data with a linear regression method \citep[the same fitting method as][]{2015EP&S...67...59M}. 
Note that the relatively shorter durations of the \textit{Hinode}'s flares are due to a selection bias that the \textit{Hinode}'s observational intervals cannot cover the overall light curves of long duration flares.   
In fact, the decay time obtained from \textit{Hinode} data is only 12 $\%$ shorter than that obtained from \textit{SDO} data (also see, Table 3).

\subsection{Comparison between Solar Flares and Superflares on Solar-type Stars} \label{subsec:33}
\begin{figure}[htbp]
\begin{center}
\includegraphics[scale=0.4]{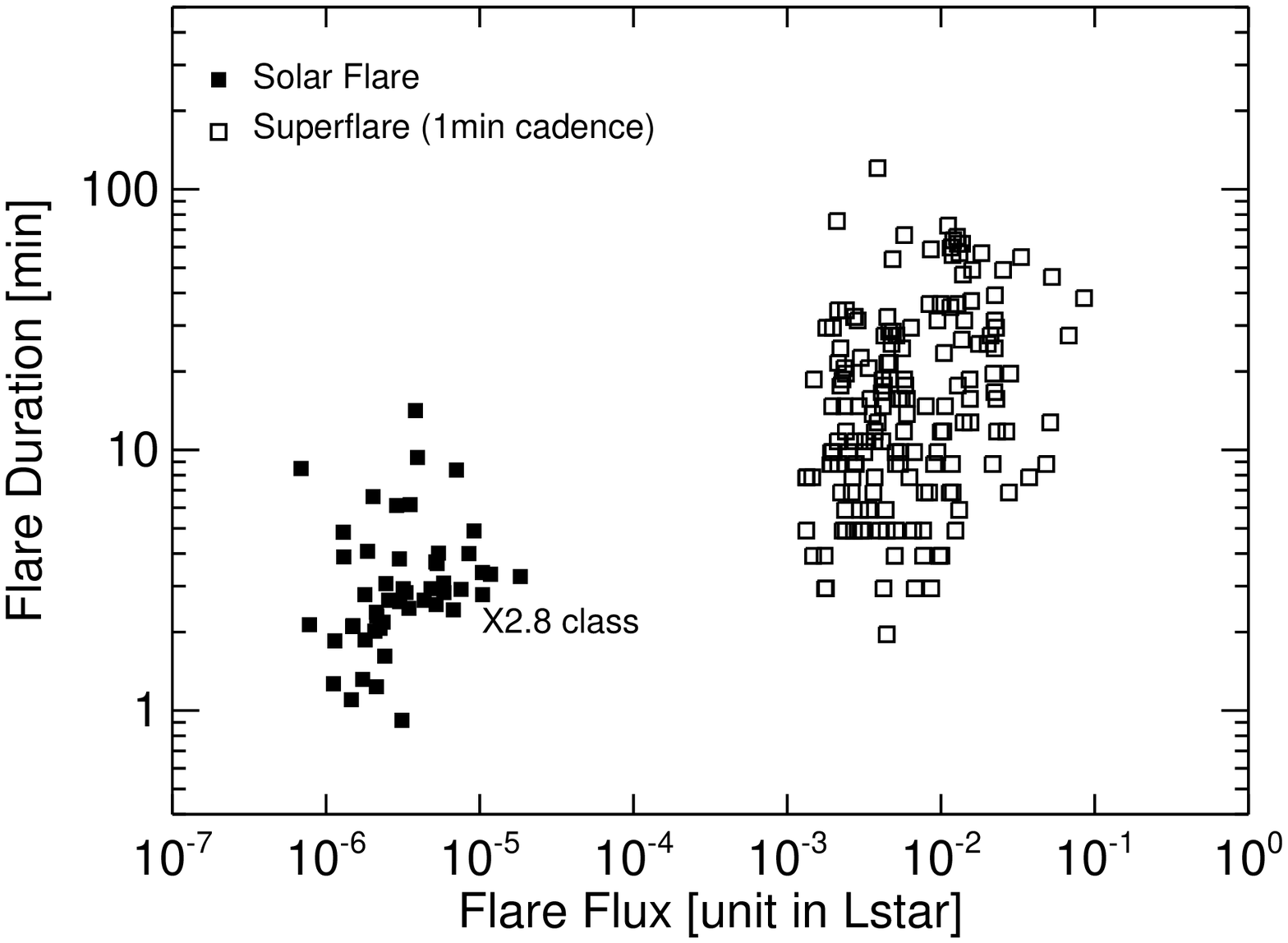}
\includegraphics[scale=0.4]{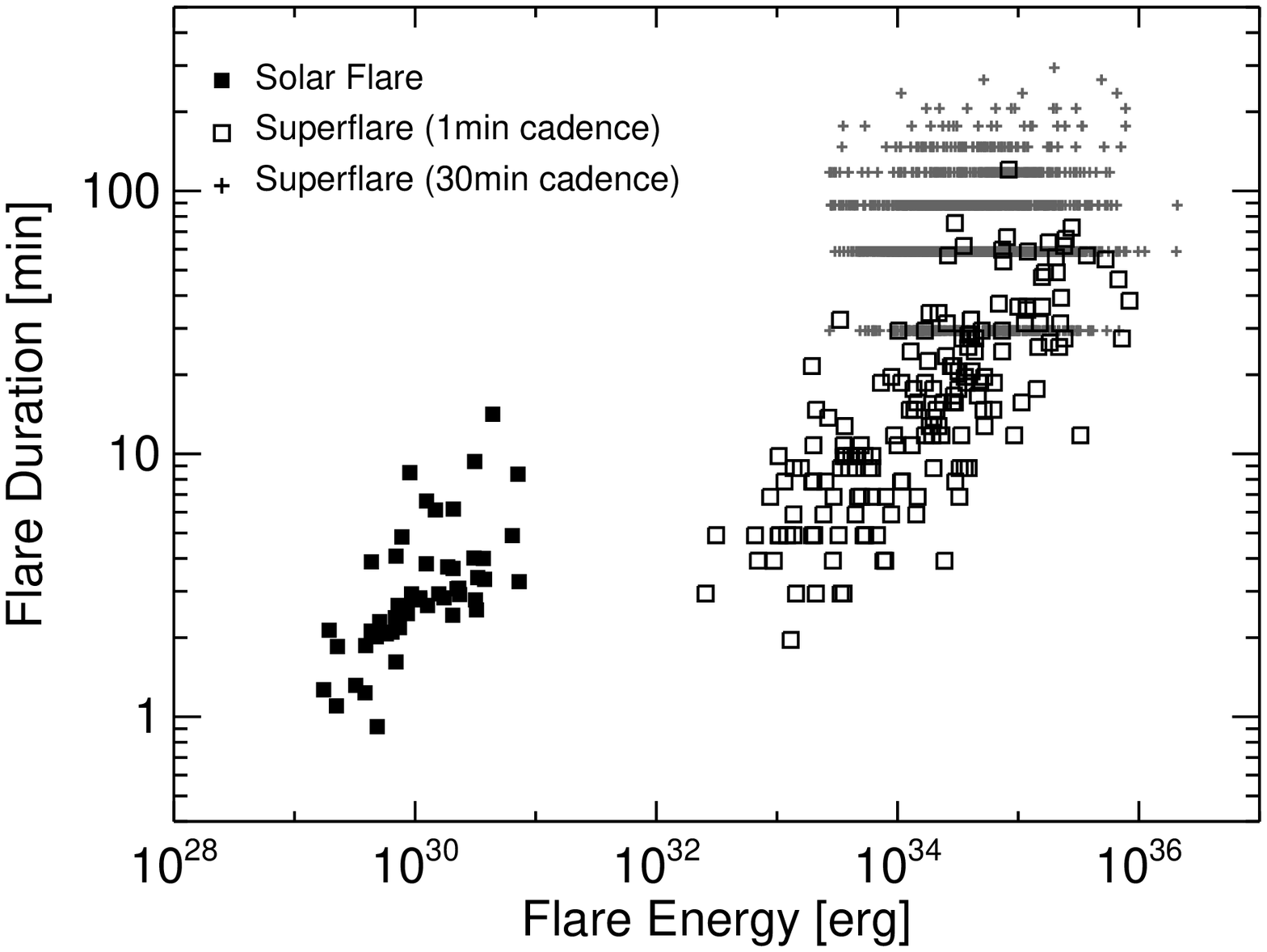}
\caption{Left: Comparison between the flare flux and duration. Right: Comparison between the flare energy and duration. The open squares show solar white-light flares analyzed in this paper and the filled squares and crosses show superflares on solar-type stars obtained from \textit{Kepler} 1 minutes and 30 minutes cadence data, respectively. The data of superflares are taken from \cite{2015EP&S...67...59M}.}
\label{fig:7}
\end{center}
\end{figure}

We compared the solar WLFs and superflares on solar-type stars in Figure \ref{fig:7}. 
The data of superflares were basically taken from \cite{2015EP&S...67...59M}, which reports 187 superflares on 23 solar-type stars.
In figure \ref{fig:7}, however, we excluded all flares (19 events in total) on KIC 7093428. 
This is because KIC 7093428 was found to be a sub-giant star (surface gravity log $g$ $\sim$ 2.77) in the latest version of the \textit{Kepler} Input Catalog \citep{2017ApJS..229...30M}, which includes the results of new estimations of log $g$ values using the granulation amplitude data in \textit{Kepler} light curve \citep[cf. Flicker method;][]{2016ApJ...818...43B}.
The other 22 superflare stars in \cite{2015EP&S...67...59M} are still in the range of solar-type stars in this revised \textit{Kepler} Input Catalog, and we use these stars in the plot in Figure \ref{fig:7}.

The left panel shows the distribution of WLF fluxes and durations of solar flares (filled squares) and superflares with short cadence data (open squares). 
The solar WLF fluxes $L_{\rm WL}$ are about $10^{-5}$--$10^{-6}L_{\rm sun}$, where $L_{\rm sun}$ is the solar luminosity. 
These values are consistent with the result of flare observations using the total solar irradiance data \citep[e.g.,][]{2011A&A...530A..84K}. 

Likewise, the right panel shows the distribution of radiated energies and durations.
The power-law index of the superflares (without KIC 7093428) is $0.38\pm 0.02$.
This is not so different from the original value (0.39), which includes the data points of KIC 7093428.
As in the above section, we found that a power-law index of solar WLFs (0.38) is found to be consistent with that of the superflares (0.38).
However, these distributions of solar and stellar flares cannot be explained by a same power-law relation \citep[cf, Equation \ref{eq:3},][]{2015EP&S...67...59M}, and the durations of superflares are one order of magnitude shorter than those extrapolated from the power-law relation of solar WLFs. 

\section{Discussion} \label{sec:dis}
\subsection{Energetics of Solar Flares} \label{subsec:41}
Many stellar flares are recently observed especially as WLFs thanks to the \textit{Kepler} data, and the properties of stellar WLFs have been studied. 
In contrast, the energy scales of solar flares are mainly classified by the \textit{GOES} X-ray flux \citep[a few $\%$ of the total radiated energy;][]{2012ApJ...759...71E}.
To compare the properties of solar and stellar flares and understand the energetics of flares, it is necessary to investigate the relation between the \textit{GOES} soft X-ray fluxes and WLF energies.

Our results of comparisons between WLF flux and \textit{GOES} soft X-ray flux ($F_{WL}\propto F_{SXR}^{0.59\pm 0.04}$) well match the relation ($F_{WL}\propto F_{SXR}^{0.65}$) obtained by \cite{2011A&A...530A..84K}. 
Even if the relation is universal across to wide energy ranges, the absence of a linear correlation is not surprising and can be explained as follows.
The relation between HXR and SXR flux is expressed as $F_{HXR}\propto dF_{SXR}/dt \propto F_{SXR}/\tau$ in the impulsive phase, which is known as Neupert effect \citep{1968ApJ...153L..59N}. 
By considering that WLFs are well correlated with HXR flares and using the relation $\tau \propto E^{1/3}$, the relation between WL fluxes and SXR fluxes is derived as $F_{WL}\propto F_{HXR} \propto F_{SXR}/\tau \propto F_{SXR}^{2/3}$, which is in agreement with the above observed $F_{WL}$--$F_{SXR}$ relations.

\begin{figure}[p]
\begin{center}
\plottwo{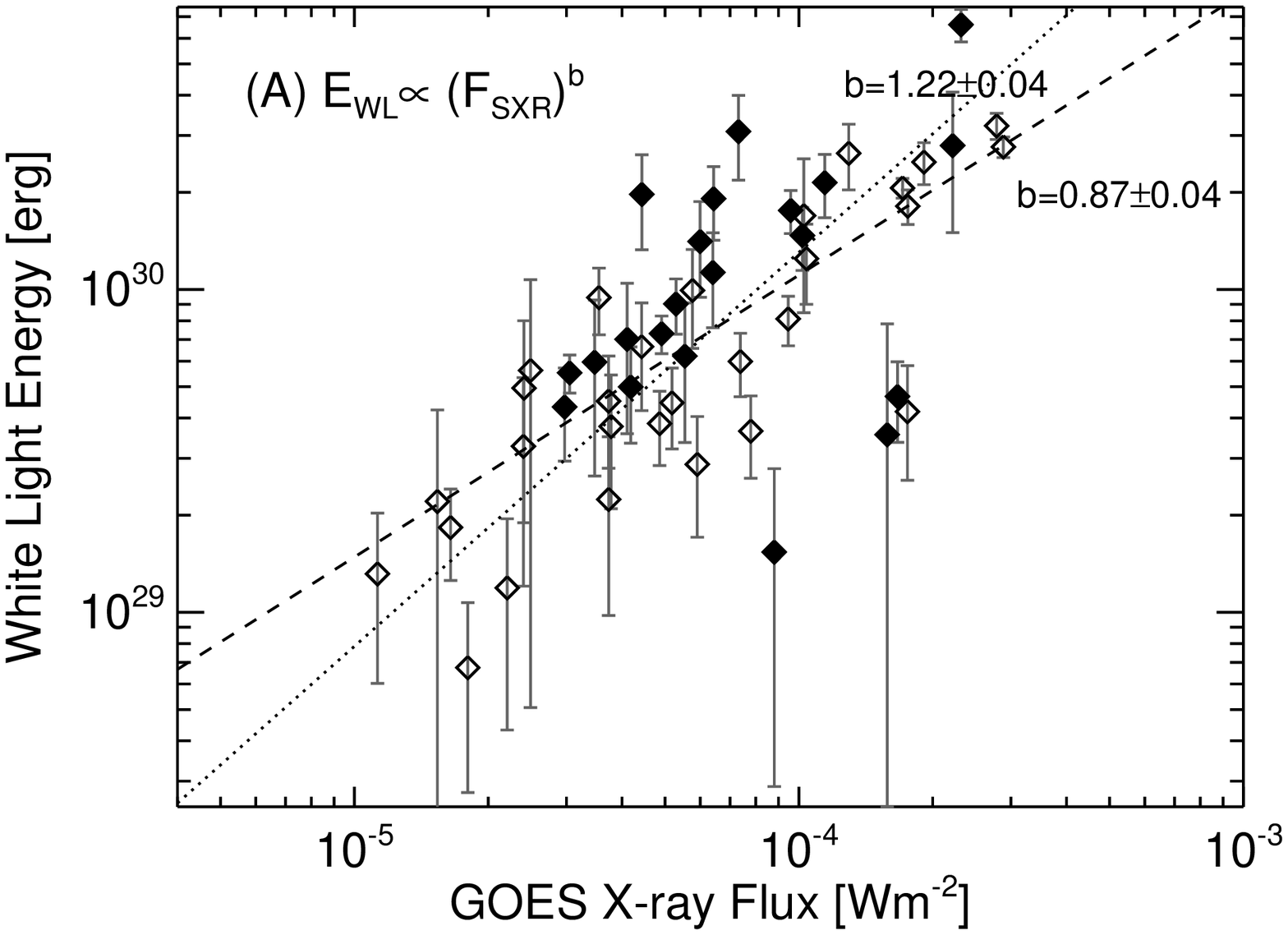}{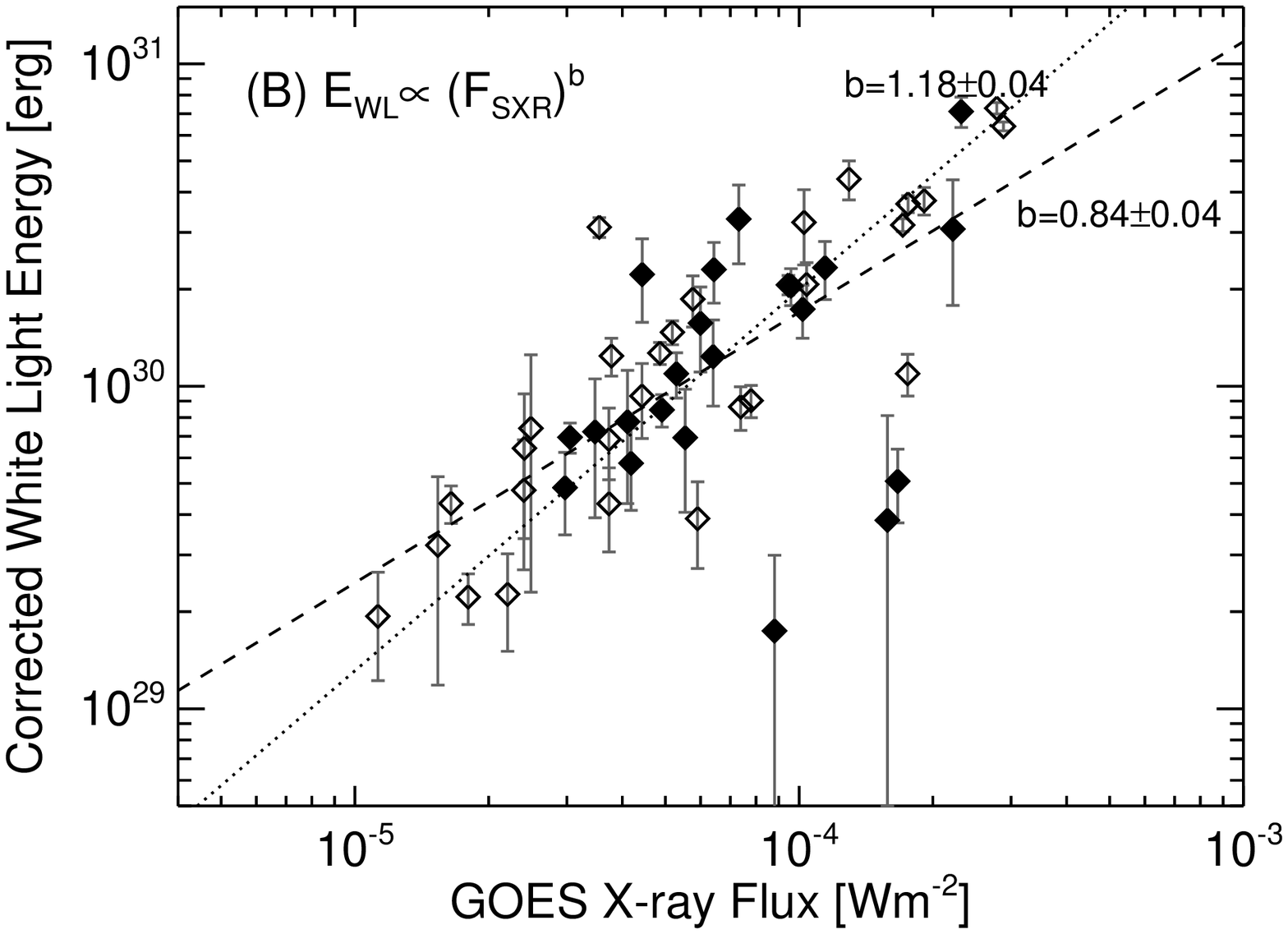}
\caption{(A) The left panel shows the comparisons of the GOES soft X-ray fluxes at the flare peak time and the HMI white-light energies. (B) The right panel shows the same as the left panel, but the white-light energies are corrected assuming the limb-darkening of the plane-parallel atmosphere. In each panel, the filled symbols are flares on the disk center whose distances from the solar center are less than 700 arcsec, and open ones are flares on the limb. Dashed and dotted lines are fitted lines with linear regression method (LR) and a linear regression bisector method \citep[LRB; ][]{1990ApJ...364..104I}, respectively. The error bars were calculated from 1$\sigma$ components of the light curve in quiescence.
The power-law relations ($E_{\rm WL}\propto (F_{\rm SXR})^{b}$) are obtained with the indexes $b=0.87\pm 0.04$ (LR) and $b=1.22\pm 0.04$ (LRB), and those of the right panel with $b=0.84\pm 0.04$ (LR) and $b=1.18\pm 0.04$ (LRB), which shows that the WL energy is proportional to \textit{GOES} soft X-ray flux.
}
\label{fig:5}
\end{center}
\end{figure}

Against these observations, \cite{2013ApJS..209....5S} suggested that the flare energy-frequency distribution cannot be explained by the Krezschmar's relation but by proportional relation ($F_{WL}\propto F_{SXR}$).
However, the relation between flare energy and soft X-ray flux can be calculated as $E_{WL}\sim F_{WL}\times \tau_{\rm dur\it} \propto F_{WL}^{3/2} \propto F_{SXR}$ by using the Krezschmar's relation and the relation of $\tau \propto E^{1/3}$.
Hence, there is in fact no contradiction between the observed flare energy-frequency distributions and the Krezschmar's relation. 
As in Figure \ref{fig:5}, we found that the relation between the flare energy and \textit{GOES} flux is in $E_{WL} \propto F_{SXR}^{0.84\pm 0.04}$ with a linear regression method.
Note that this power-law index may not be universal among wide energy ranges because it significantly depends on the fitting method due to such a narrow x-y range \citep[e.g, $E_{WL} \propto F_{SXR}^{1.18\pm 0.04}$ with a linear regression bisector method;][]{1990ApJ...364..104I}. 

\subsection{The Energy and Duration Diagram} \label{subsec:42}
\begin{figure}[htbp]
\begin{center}
\includegraphics[scale=0.5]{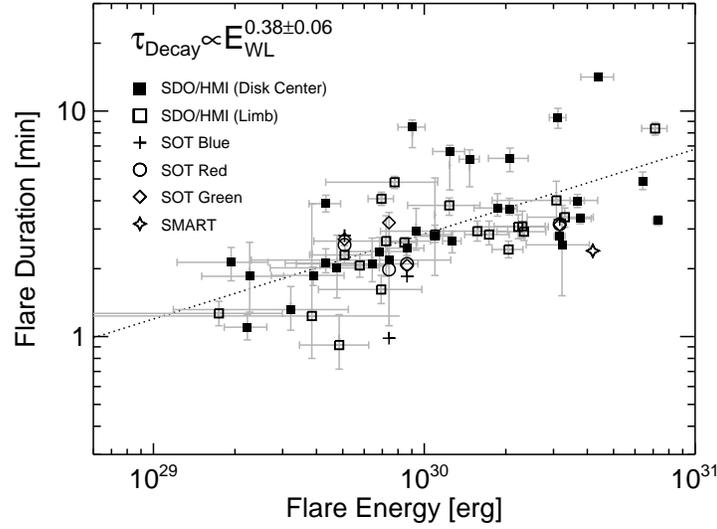}
\caption{Comparison between the flare energy and duration. The squares show solar white-light flares analyzed in this paper and the dotted line is a fitting result for these data with a linear regression method. The filled squares are flares on the disk center whose distances from the solar center are less than 700 arcsec, and open ones are flares on the limb. The data whose durations were observed by \textit{Hinode} SOT blue, red and green continuum were also plotted with crossed, open circles and open diamonds, respectively. The data observed by \textit{SMART} is also plotted with a star symbol.}
\label{fig:6}
\end{center}
\end{figure}

As mentioned in Section \ref{sec:int}, the relations between flare energies and durations have been found to be universal among solar X-ray flares ($\tau \propto E^{0.2-0.33}$) and stellar WLFs ($\tau \propto E^{0.39}$), and the relation well matches the theoretical relation consistent with magnetic reconnection ($\tau \propto E^{1/3}$).
Our result also showed that the relation of solar WLFs ($\tau \propto E^{0.38\pm 0.06}$) well matches these previous studies (Figure \ref{fig:6}). 
This consistency supports the suggestion that both solar and stellar flares are caused by the magnetic reconnection. 
However, as in Figure \ref{fig:7}, it was also found that WLFs on the Sun and the solar-type stars were not on a same line though the power-law indexes are the same ($\tau \propto E^{1/3}$).
This discrepancy indicates that solar and stellar WLFs cannot be simply explained by the relation derived by \cite{2015EP&S...67...59M}.
We propose two possibilities to explain such a discrepancy:
\begin{itemize}
\item properties of cooling or heating mechanisms unique to WLFs (Section \ref{dis:cause3}).
\item differences in physical parameters between solar and stellar flares (Section \ref{dis:cause2})
\end{itemize}
Of course some analysis problems cannot be completely excluded. 
We will therefore discuss the validity of our analyses in Appendix \ref{app1}.
Nevertheless, the result can have a potential to know the difference between solar flares and superflares as well as the unsolved mechanism of WLFs.
In the following sections, we discuss the above two possibilities in detail.

\subsubsection{Properties of Cooling or Heating Mechanisms of WLFs}\label{subsec:423}\label{dis:cause3}

The first interpretation of the $E$-$\tau$ diagram is that the difference between solar and stellar flares is related to the properties unique to WLFs, especially cooling effect.
It is observationally known that solar WLFs have two emission components:
a ``core'' structure which is a candidate of direct chromospheric heating, 
and a ``halo'' structure which is a candidate of backwarming \citep{2007PASJ...59S.807I}.
High-time-resolution observations also reveal that, in the ``halo'' structure, there are long ($\sim$500 sec) decay components and the timescales correspond to coronal cooling timescales \citep{2016ApJ...833...50K}.
We should note that it is not confirmed whether such decay time originates in the cooling timescale or the long lasting reconnection.
However, if it does correspond to the cooling timescale, the 500 sec cooling time ($t_{\rm cool}$) is not negligible compared to the reconnection timescale ($t_{\rm rec}$) and could elongate the decay time of WLFs.
Therefore, there is a possibility that the decay time of solar WLFs (below $\sim10^{32}$ erg) are elongated by cooling effect because $t_{\rm cool} >> \it t_{\rm rec}$, but those of superflares (above $\sim10^{33}$ erg) are not elongated because $t_{\rm cool} << \it t_{\rm rec}$, which would result in the observed $E$--$\tau$ discrepancy.

\begin{figure}[htbp]
\begin{center}
\includegraphics[scale=0.5]{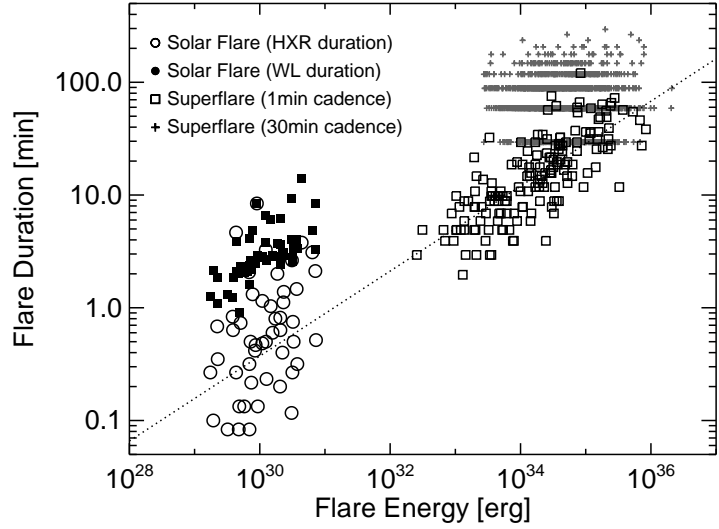}
\caption{
The symbols are basically the same as those in right panel of Figure \ref{fig:7}, but open circles are solar flares whose durations are replaced for those of HXR flares.
The HXR durations are defined as those from the peak time to the maximum time whose intensity are over 1/e of the peak values.
}
\label{fig:9.5}
\end{center}
\end{figure}

We consider that comparisons of the durations of solar WLFs and HXR flares can provide us a hint to cooling effect because the HXR nonthermal emissions are not affected by cooling time.
We then measured the decay time of HXR emissions (30--80 keV) as a proxy of the reconnection timescale by using \textit{RHESSI} data.
In Figure \ref{fig:9.5}, open circles are solar flares whose durations are replaced for those of HXR flares.
The durations of solar HXR flares are shorter by a factor of 5 at average than those of WLFs.
This indicates that the durations of solar WLFs are somewhat elongated by any cooling effect.
Also, Figure \ref{fig:9.5} shows that the distribution is roughly on the line extrapolated from the distribution of the superflares.
If we could assume that $t_{\rm cool} << t_{\rm rec}$ in the case of superflares, then we might explain the observed $E$--$\tau$ diagram and conclude that both solar flares and stellar superflares are explained by the magnetic reconnection theory.

Note that we cannot completely support the above suggestion due to the following reasons.
First, we have no knowledge about the relation between HXR durations and WL ones of superflares.
Therefore, we cannot exclude a possibility that WL emissions of stellar superflares are also affected by cooling effect.
Secondly, as in Figure \ref{fig:7}, the $E$-$\tau$ relation of solar WLFs can be explained by the same theoretical line ($\tau \propto E^{1/3}$) as the large part of long-duration superflares observed by Kepler 30 min cadence.
This may indicate that the WL cooling effect is not enough to explain the $E$--$\tau$ diagram.

In this section, we examined the possibility that the cooling timescale might be important in the understandings of the observed $E$-$\tau$ diagram.
This may resolve the observed discrepancy except for some problems.
To conclude how essential the cooling effect is, we have to (i) reveal the emission mechanism of solar WLFs, or to (ii) carry out multi-wavelength observations (WL and HXR) of superflares.

\subsubsection{Difference in Physical Parameters between Solar and Stellar Flares}\label{dis:cause2}
We present here the second interpretation of the observed $E$--$\tau$ diagram.
\cite{2015EP&S...67...59M} derived the theoretical scaling law $\tau \propto E^{1/3}$ assuming that the Alfv\'{e}n velocity ($v_A=B/\sqrt{4\pi\rho}$) around reconnection region is constant among each flare on solar-type stars.
This assumption would be roughly appropriate according to solar and stellar observations \citep[e.g.,][]{1999ApJ...526L..49S,2002ApJ...577..422S}. 
When considering the dependence on Alfv\'{e}n velocity, the scaling law can be expressed as follows:
\begin{eqnarray}\label{eq:mae}
\tau \propto E^{1/3}B^{-5/3}\rho^{1/2}.
\end{eqnarray} 
On the basis of this scaling law, the one order of magnitude shorter durations of superflares can be understood by (1) two orders of magnitude lower coronal density of superflares, or (2) about a factor of 2--4 stronger coronal magnetic field strength of superflares than that of solar flares. 
The former possibility is less likely because superflare stars are rapidly rotating ones which are expected to have higher coronal densities based on the large emission measures of the X-ray intensity \citep[e.g.,][]{2011ApJ...743...48W}.
On the other hand, the latter well accounts for the $E$--$\tau$ distributions without any contradiction with observations that superflare stars show high magnetic activities \citep[e.g.,][]{2015PASJ...67...33N}.
On the basis of such a scaling relation, we proposed that the discrepancy can be caused by the strong coronal magnetic field strength of superflares.

\begin{figure}[htbp]
\begin{center}
\includegraphics[scale=0.5]{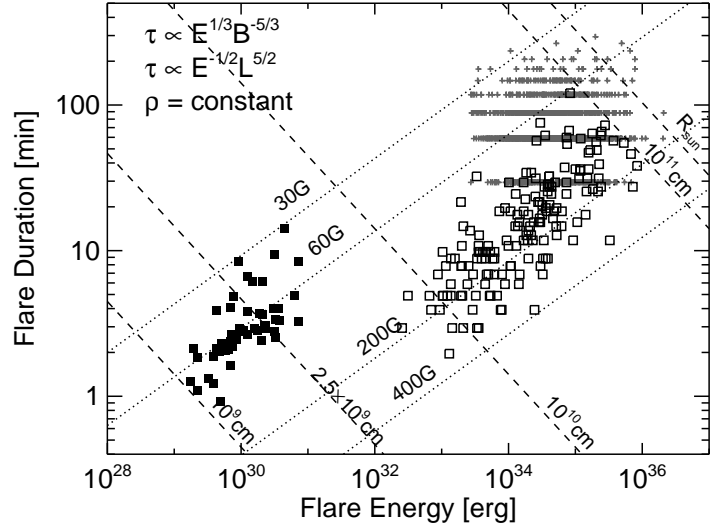}
\caption{Theoretical $E$--$\tau$ relations (Equation \ref{eq:nam}, \ref{eq:nam2}) overlaid on the observed $E$--$\tau$ relation in this study and Maehara et al. (2015). Theoretical lines are plotted with dotted lines for the different magnetic field strength $B$ = 30, 60 (an observational value of the solar flares), 200 and 400 G. The flare loop length $L$ = constant lines are also plotted with dashed lines for $L$=$10^9$, $2.5\times 10^9$ (an observational), $10^{10}$, $10^{11}$ cm and a solar diameter.}
\label{fig:8}
\end{center}
\end{figure}

By assuming pre-flare coronal density is a constant value, Equations \ref{eq:1} and \ref{eq:2} give the following new scaling laws:
\begin{eqnarray}\label{eq:nam}
\tau \propto E^{1/3}B^{-5/3}\\
\tau \propto E^{-1/2}L^{5/2}.\label{eq:nam2}
\end{eqnarray}
To simply determine the coefficients, we observationally measured the average values of $B$ and $L$ on the basis of the method introduced by \cite{2017PASJ...69....7N}.
$B$ is extrapolated from the photospheric magnetic fields using \textit{SDO}/HMI magnetogram, and $L$ is calculated as square roots of the flaring area observed with \textit{SDO}/Atmospheric Imaging Assembly \citep[AIA;][]{2012SoPh..275...17L} 94 {\AA} (see in detail Appendix \ref{sec:43}).
As a result, the coefficients can be obtained as $B_0=57$ G, $L_0=2.4\times 10^9$ cm, $\tau_0=3.5$ min and $E_0=1.5\times10^{30}$ erg from the average among the solar flares in our catalogue.
On the basis of such values, we applied the scaling laws to the observed $E$--$\tau$ diagram as in Figure \ref{fig:8}, and found that solar flares and stellar flares have coronal magnetic field strength of 30--400 G.
This range is roughly comparable to the observed values (40--300 G) of solar and stellar flares \citep{1973SoPh...33..445R,1985ARA&A..23..169D,1996ApJ...456..840T,1997Natur.387...56G} and those (15--150 G) estimated by \cite{1999ApJ...526L..49S,2002ApJ...577..422S}.
Moreover, it is reasonable that the estimated loop length of superflares ($\sim 10^{10}$--$10^{11}$ cm) is less than the solar diameter ($1.4\times10^{11}$ cm).
These consistencies support our suggestions.
According to the scaling law, superflares observed with short time cadence have 2--4 times stronger coronal magnetic field strength than solar flares.
Although it would be controversial whether such strong coronal magnetic fields can be really formed and sustained in the stellar coronae,
it may be possible that, in the case of the superflare stars, the stronger fields of the surrounding quiet regions or the large star spots suppress the magnetic loops of active regions, and then sustain such strong magnetic fields. 
Note that this does not mean that all of stellar flares can be caused in such strong magnetic fields.
As seen in the left and right panels of Figure \ref{fig:7}, the upper limit of $E$-$\tau$ diagram is determined by the detection limit of superflares, implying a selection bias that the detection method by \cite{2015EP&S...67...59M} tends to select relatively impulsive superflares. 
Therefore, it is natural that there are expected to be long-duration superflares with field strengths of a few 10 G .

\begin{figure}[htbp]
\begin{center}
\includegraphics[scale=0.42]{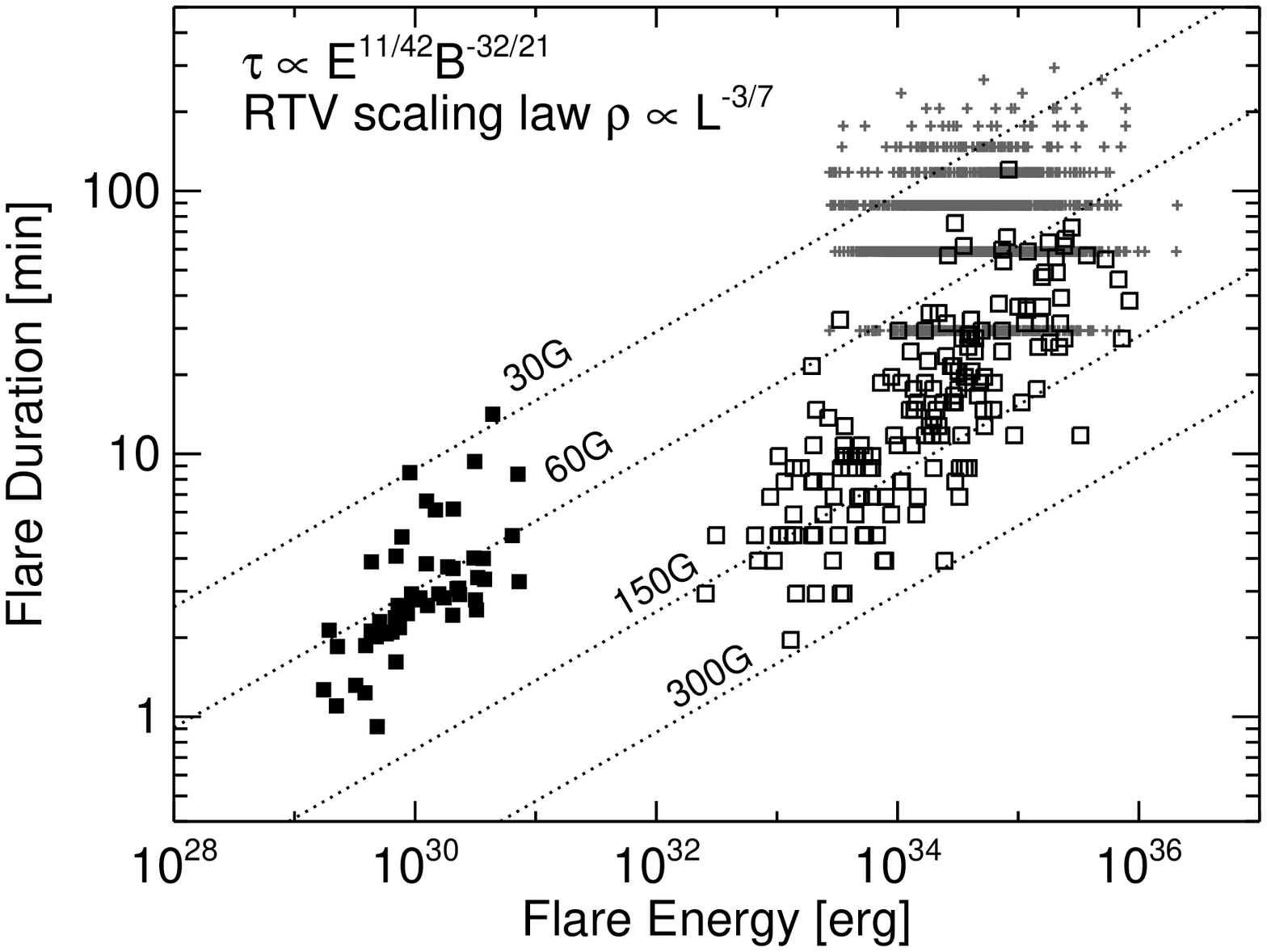}
\includegraphics[scale=0.42]{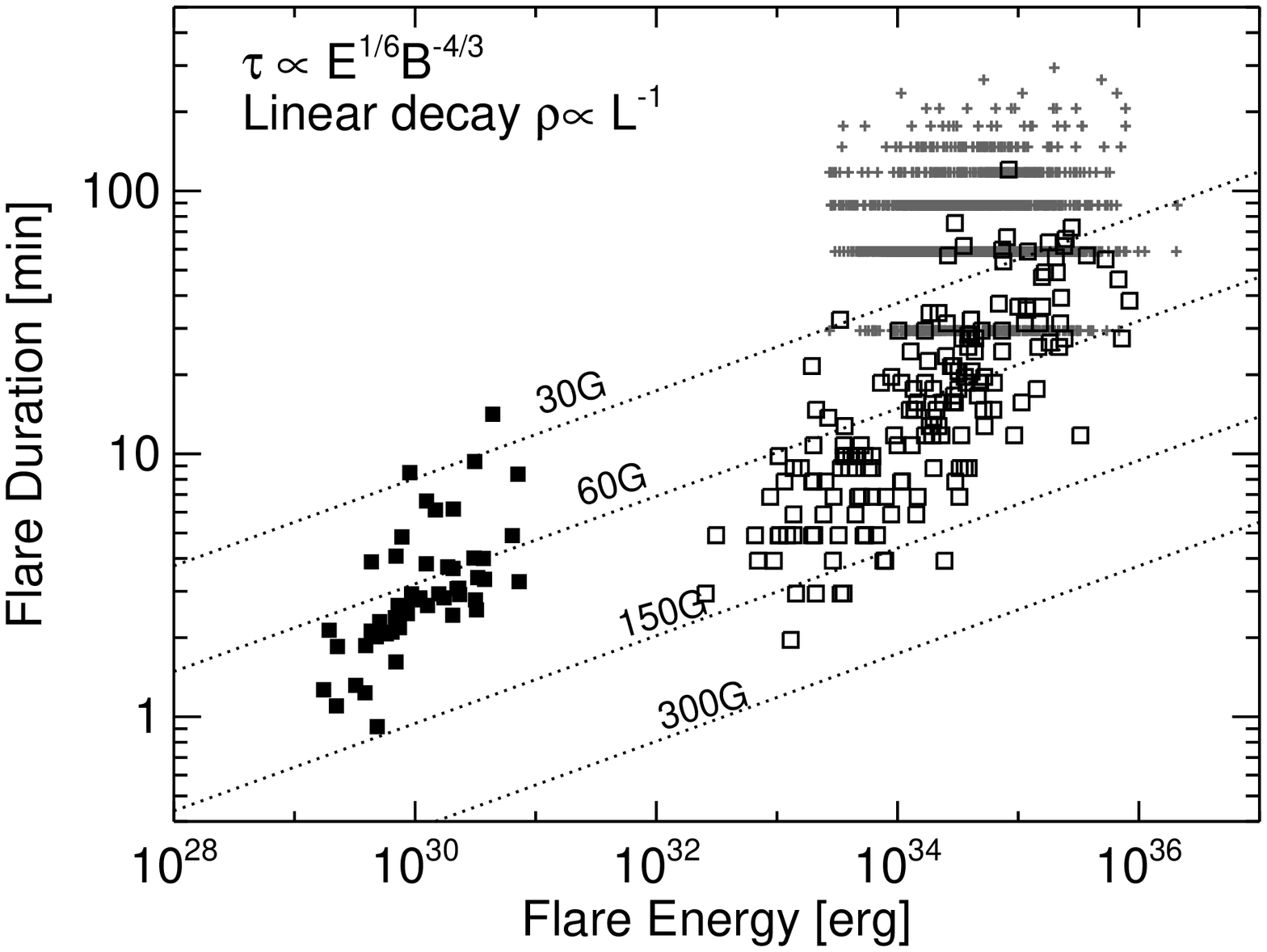}
\caption{
The figure shows the theoretical $E$--$\tau$ relations (Equation \ref{eq:rtv}) considering the dependence of the pre-flare coronal density, as well as the observed data. Four theoretical lines are plotted for the different magnetic field strength $B$ = 30, 60, 150 and 300 G.
}
\label{fig:9}
\end{center}
\end{figure}

To explain the observed $E$-$\tau$ diagram more realistically, we incorporated the dependence of the pre-flare coronal density ($\rho$) into the scaling law in the following two ways: 
(1) RTV (Rosner--Tucker--Vaiana) scaling law $\rho \propto L^{-3/7}$ (2) gravity stratification $\rho \propto L^{-1}$.
RTV scaling law \citep[$T\propto (pL)^{1/3}$; $T$ is temperature, $p$ is gas pressure; ][]{1978ApJ...220..643R} are well adopted to the static magnetic loop on the solar surface and can be deformed to $\rho \propto F^{4/7}L^{-3/7}$, where $F$ is coronal heating flux.
In the case of gravity stratification, the coronal density linearly decreases as the height increases by a simple approximation \citep{2016ApJ...833L...8T}.
In both cases, a pre-flare coronal density can be expressed as $\rho \propto L^{-a}$ ($\rm a>0$).
When the density dependence is considered, the scaling relation is written as:
\begin{eqnarray}\label{eq:rtv}
\tau \propto E^{1/3-a/6}B^{-5/3+a/3}.
\end{eqnarray} 
Although we ignored the $F$ dependence, it would be negligible due to the weak dependence of $F^{2/7}$. 
In Figure \ref{fig:9} , the theoretical scaling laws are plotted as well as the observed data. 
We found that the scaling law can explain the solar and stellar observation with roughly the same magnetic field strength (30--300 G). 
The power-law distributions of solar and stellar flares imply that the magnetic field strengths decrease as the magnetic loops expand, and the different distributions of solar and stellar flares can also be explained by the different magnetic field strength at a given height. 
In the case of comparison among different spectral types, the problem would become complex because the $F$ dependence cannot be negligible. 
The validity of the scaling law will be examined in Appendix \ref{sec:43},
and the comparison between the different spectral type will be discussed in Appendix \ref{app4}.

In this section, we derived the scaling laws on the basis of magnetic reconnection theory, and suggested that the observed $E$--$\tau$ distribution can be explained by the different magnetic field strength of solar and stellar flares, regardless of the WL cooling effect.
These discussions in turn imply that the stellar properties can be estimated by using the scaling law from the most simply observable physical quantities (flare energy and duration).
This would be helpful for researches on stellar properties in the future photometric observations (e.g., $TESS$, \citealt{2015JATIS...1a4003R}).
It is, however, not clear that the scaling law can be applicable to the real observations due to the lack of the validation.
As a future study, the measurements of magnetic field of superflare stars would be necessary to ascertain whether the observed $E$--$\tau$ diagram are caused by magnetic field strength or cooling effect.


\section{Summary} \label{sec:sum}

We conducted a statistical research on solar WLFs, and compared the relation of flare energy ($E$) and duration ($\tau$) with those of superflares on solar-type stars, aiming to understand the energy release of superflares by the magnetic reconnection theory. 
The results show that superflares have one order of magnitude shorter durations than those extrapolated from the power-law relation of the obtained solar WLFs.
This discrepancy may have a potential to understand the detailed energy release mechanism of superflares as well as properties of the unsolved origin of WL emissions.
To explain this result, we proposed the following two physical interpretations on the $E$--$\tau$ diagram.

\begin{enumerate}
\item In the case of solar flares, the reconnection timescale is shorter than the cooling timescale of white light, and the decay time is determined by the cooling timescale.
\item The distribution can be understood by a scaling law ($\tau \propto E^{1/3}B^{-5/3}$) obtained from the magnetic reconnection, and the coronal magnetic fields of the observed superflares are 2-4 times stronger than those of solar flares.
The scaling laws can predict the unresolved stellar parameters, the magnetic field strength and loop length.
This would be helpful for investigations on stellar properties in the future photometric observations (e.g., $TESS$).
\end{enumerate}
However, both cases are not enough validated.
In case (1), the lack of our understanding of WLFs prevents us from accepting it.
It is therefore necessary to reveal the emission mechanism or to detect the hard X-ray emission of superflares to examine the effect of the cooling timescale.
On the other hand, in case (2), the validation for solar flares are discussed in Appendix \ref{sec:43}, but it has no clear evidence due to its complexity.
We expect that measurements of stellar magnetic field strengths by other methods will give us the answer.

\bigskip
Acknowledgement: We acknowledge with thanks P. Heinzel, T. Kawate, S. Masuda, S. Takasao, and T. Takahashi for their contribution of fruitful comments on our work.
We also thank S. Hawley and J. Davenport for kindly providing us the data of their papers.
\textit{SDO} is part of NASA’s Living with a Star Program. 
\textit{RHESSI} is the NASA Small Explorer mission.
\textit{Kepler} was selected as the 10th Discovery Mission.
\textit{Hinode} is a Japanese mission developed and launched by ISAS/JAXA, with NAOJ as domestic partner and NASA and STFC (UK) as international partners.
This work was supported by JSPS KAKENHI Grant Numbers 
JP26800096, 
JP26400231, 
JP15K17772, 
JP15K17622, 
JP16H03955, 
JP16H01187, 
JP16J00320, 
JP16J06887, 
JP17H02865, 
and JP17K05400. 

\clearpage
\clearpage
 
\clearpage
\clearpage
 
\clearpage


\appendix
\section{The Validity of Our Analyses}\label{app1}
Here, we examined the validity of our analysis on the basis of the following three points: (1) the assumption of 10,000 K blackbody radiation, (2) the relatively low time cadence of \textit{SDO}/HMI, and (3) the integration regions of WL emission.

\begin{figure}[p]
\begin{center}
\includegraphics[scale=0.5]{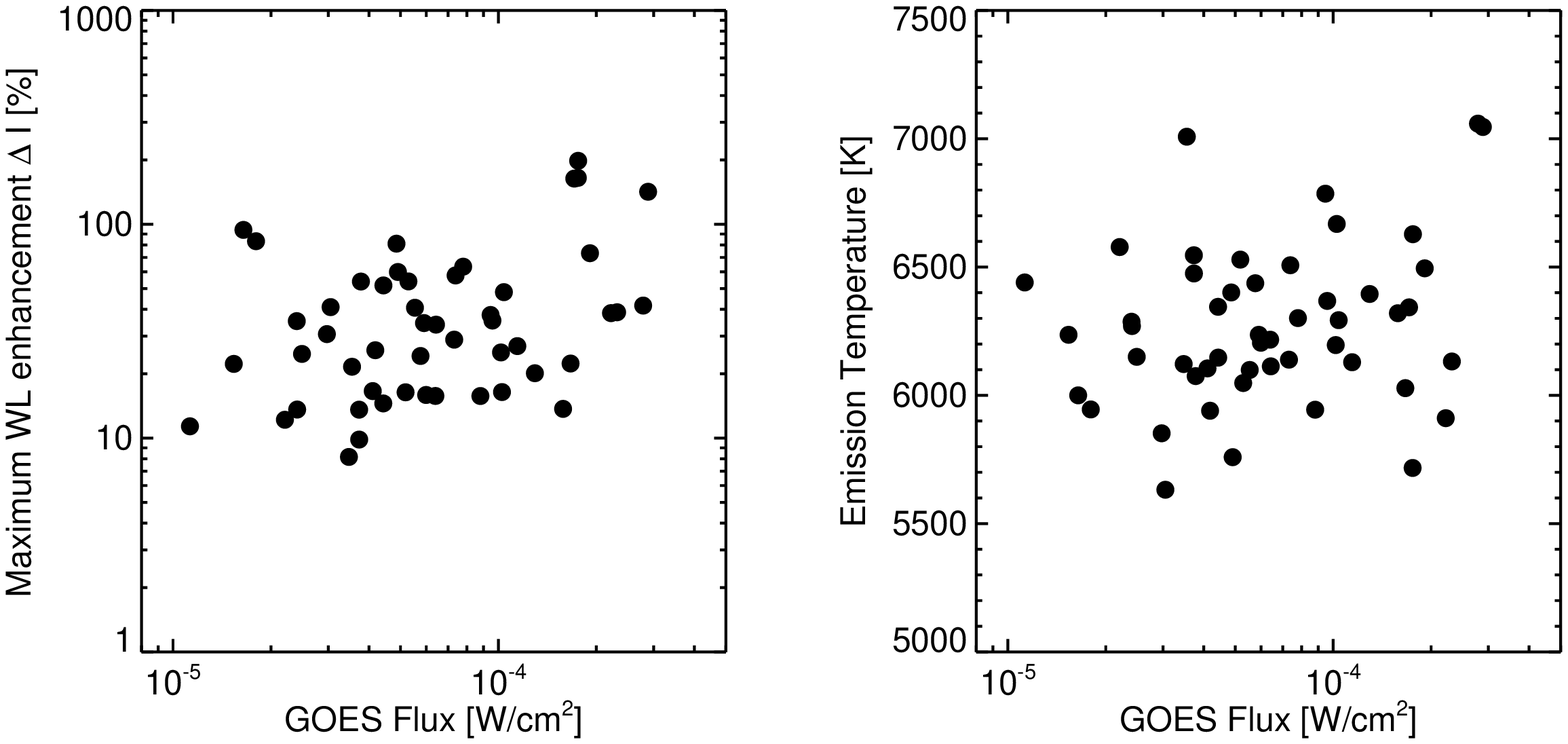}
\caption{The left panel shows comparisons between \textit{GOES} X-ray flux and maximum WL enhancements compared to the pre-flare levels. We used the \textit{SDO}/HMI data at the flare peaks for our catalogue. The right panel is comparisons of \textit{GOES} X-ray flux and the emission temperature of the WL enhancements calculated by the Equation \ref{eq:temp}.}
\label{fig:ap1}
\end{center}
\end{figure}

The problem (1) may be one of the most doubtful points, but we show that it is not essential in the case of our analysis.
\cite{2015EP&S...67...59M} assumed the blackbody radiation 10,000 K on the basis of the ``blackbody like'' spectra of M-type star flares \citep[e.g., ][]{1992ApJS...78..565H} and a few observations of solar WLFs \citep[e.g., ][]{2011A&A...530A..84K}. 
Here we have to mention the remark of \cite{2016ApJ...816...88K} that the temperature measurements of \cite{2011A&A...530A..84K} might be somewhat inaccurate because the temperature is obtained by assuming optically thin radiation. 
When assuming an optically thick radiation, the temperature decreases by an order of 1,000K.
In fact, observations using \textit{Hinode} and \textit{SDO} show that the radiation temperature is $\sim$6000 K of WLFs \citep[e.g., ][]{2013ApJ...776..123W,2014ApJ...783...98K,2016ApJ...816...88K}.
We also carried out a pixel-based analysis by using \textit{SDO}/HMI data, revealing that the enhancement of WLFs are typically 10-50\% and the emission temperature of the brightest pixel is about 5500--7000K as in Figure \ref{fig:ap1}.
The left panel of Figure \ref{fig:ap1} shows a comparison between \textit{GOES} X-ray flux and maximum WL enhancements ($\Delta I =I_{\rm flare}/I_{\rm quiescence}$) of \textit{SDO}/HMI at the flare peaks in our WLF catalogue.
By assuming a blackbody radiation with single temperature, we calculated the WL emission temperature ($T_{\rm WL}$) of the pixels on the basis of emission level ($I_{\rm QR}$) of quite region where temperature ($T_{\rm eff}$) is $\sim$5800K:
\begin{eqnarray}\label{eq:temp}
I_{\rm WL}/I_{\rm QR} &=& B(T_{\rm WL}; \lambda= \rm 6173\AA\it)/B(T_{\rm eff}; \lambda=\rm 6173\AA\it),
\end{eqnarray}
where $B$($T$,$\lambda$) is a \textit{Planck} function.
The right panel of Figure \ref{fig:ap1} is a comparison \textit{GOES} X-ray flux and the maximum emission temperature of the WL enhancements over 3$\sigma$.
Some of the emission temperatures are less than the effective temperature of the Sun ($\sim$5800K) because the large WL enhancements are often detected in the sunspots.
As one can see, the temperatures are from 5500K to 7000K, which is comparable to the previous studies based of multi-wavelength observation with \textit{Hinode} \citep[e.g., ][]{2013ApJ...776..123W}.
Therefore, the 10,000 K radiation temperature is not so well justified even for solar WLFs.
Nevertheless, in the case of our analysis, we can conclude that the difference in the emission temperature hardly affect the estimation of the energies of WLFs.
This is because the decrease in the temperature leads to the decrease in surface luminosity ($\propto T^4$), but also to the increase in emission area ($\propto T^{-4}$) when using Equation \ref{eq:ene} and \ref{eq:are}.
For example, when the temperature changes to 6000K (7000K) from 10,000 K, the estimated energies will change only by a factor of 0.49 (0.56). 
This cannot explain the gap between solar and stellar flares as far as we assume blackbody, which indicates that the assumption is not essential in this study.

\begin{figure}[p]
\begin{center}
\includegraphics[scale=0.5]{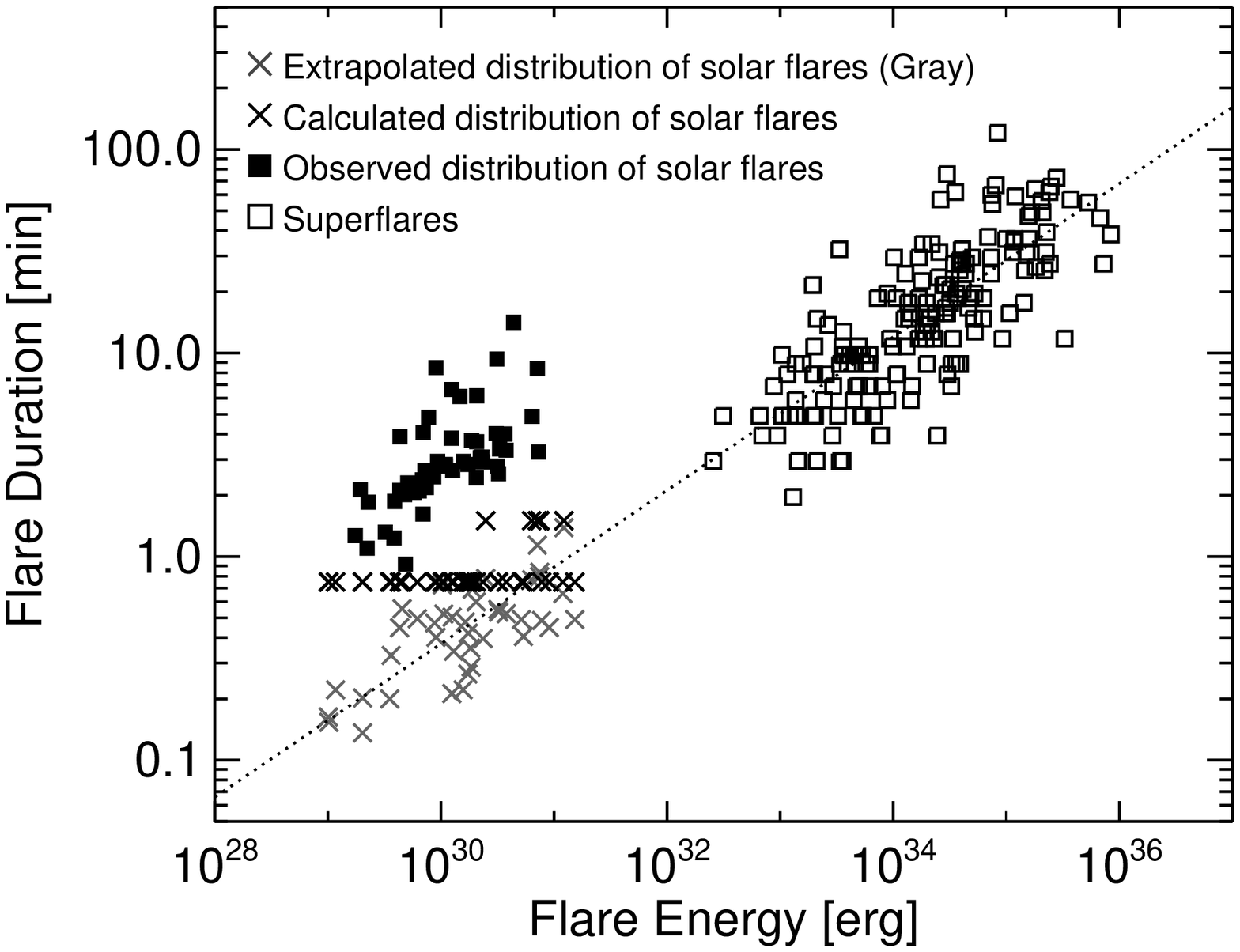}
\caption{The energy and duration diagram. The symbols are basically the same as Figure \ref{fig:7}. Dotted lines are extrapolated lines from the distributions of superflares. Gray crosses are distributions in the range of $10^{29-31}$ erg which extrapolated from those of superflares, and black ones are the calculated distributions by assuming the time cadence of 45 seconds.}
\label{fig:ap2}
\end{center}
\end{figure}

The next point (2) is that the decay time may be overestimated because the time cadence of HMI is too long to follow the temporal evolutions of solar WLFs. 
Against this problem, we confirmed that there actually exist solar WLFs having such a long duration in Section \ref{subsec:23} by comparing with the higher time-cadence observations with \textit{Hinode}/SOT and \textit{SMART}/T3. 
On the other hand, the existence of solar WLFs with ultra-short duration ($\sim$ a few sec) cannot be completely excluded.
Flares with no WL emission (non-WLFs) may be candidates of ultra-impulsive WLFs, which would be less likely because non-WLFs are typically long duration flares \citep{2017arXiv171009531W}.
To show more evidence, we demonstrate the effect of time cadence of the \textit{SDO}/HMI on the detectable durations of WLFs by simply simulating light curves as follows.
First, we assumed light curves with rapid increases of the emission and single exponential decay (i.e., $\propto \rm  exp(-\it t/t_{\rm dur})$, $t_{\rm dur} \propto E^{1/3}$).
The e-folding decay times ($t_{\rm dur}$) are extrapolated from the distribution of superflares \citep[see a dotted line in Figure \ref{fig:ap2};][]{2015EP&S...67...59M}.
Next, we calculated the e-folding decay time when an instrument has a time cadence of 45 seconds and a exposure time of 0.0025 sec.
Note that the real observation is more complex because the \textit{SDO}/HMI continuum is a ``reconstruction'' from the six filtergrams.
Figure \ref{fig:ap2} shows the result of the above analyses where the gray crosses are extrapolated distributions of solar WLFs ($t_{\rm dur}$) and the black are the calculated observable distributions ($t_{\rm obs}$).
The observed distribution of solar WLFs (filled square) cannot be explained by the calculated distribution ($t_{\rm obs}$), which indicates that the observed distribution of solar flares are not caused by the lack of the time cadence.
According to such observations and calculations, there is no clear evidence that the durations of solar WLFs are overestimated due to the time cadence of the \textit{SDO}/HMI.

\begin{figure}[htbp]
\begin{center}
\includegraphics[scale=0.5]{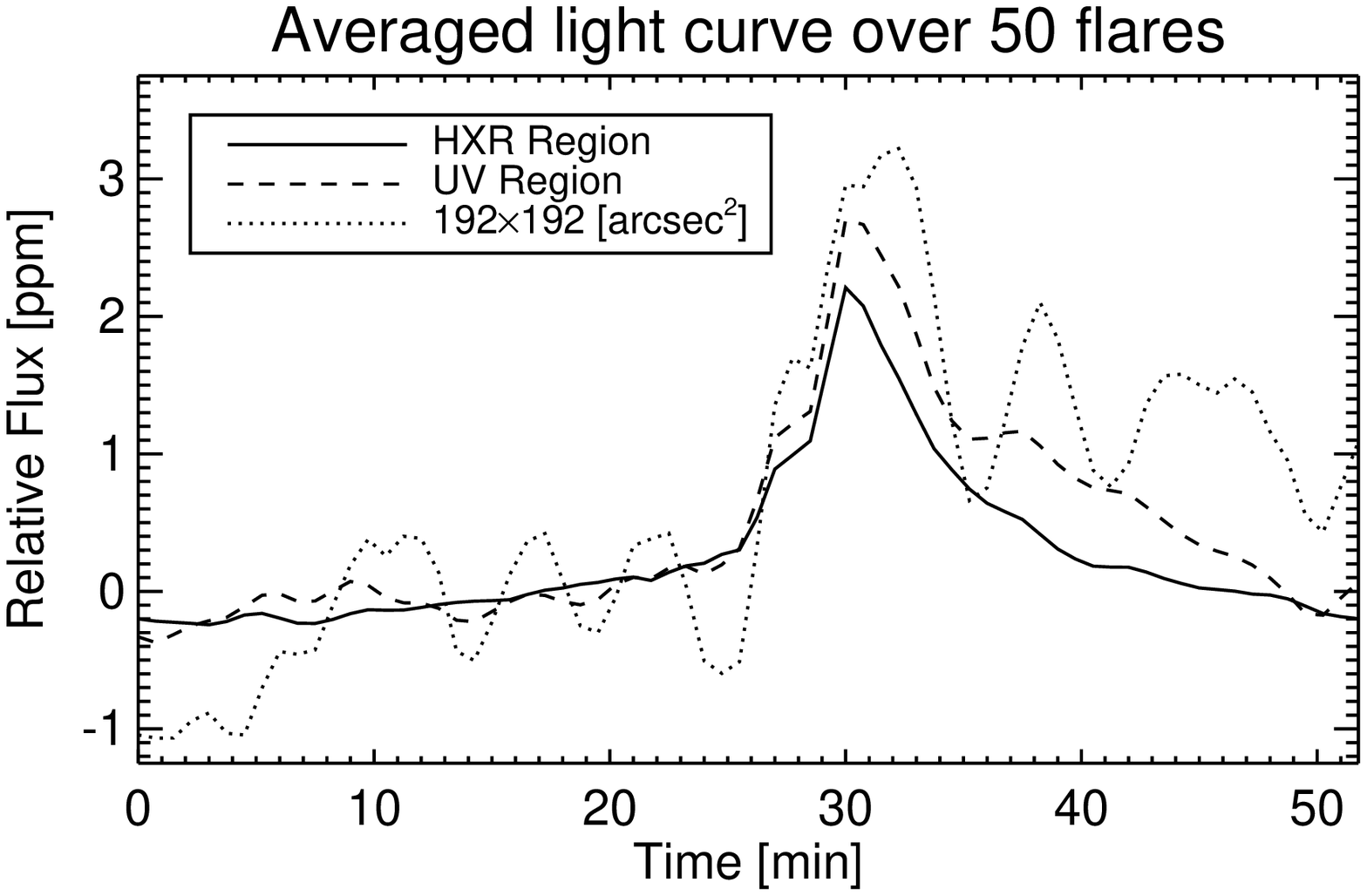}
\caption{Figure shows the light curves averaged over 50 WLFs by changing the integration region to hard X-ray region (solid), UV region (\textit{SDO}/AIA 1600 \AA; dashed) and global region (192$\times$192 arcsec around the event location; dotted).}
\label{fig:ap0}
\end{center}
\end{figure}

Lastly, we confirmed the point (3).
The integration regions were defined by the \textit{RHESSI} HXR emissions and the area was adjusted so that most of the WL emission over its 1$\sigma$ level are located inside the area.
However, there may be weak WL emission under the 1$\sigma$ level outside the HXR region.
We then measured light curves averaged over 50 WLFs by changing the integration region to UV regions (\textit{SDO}/AIA 1600 \AA) and global regions (192$\times$192 arcsec around the event location), which are much larger than HXR compact source.
Here we used the AIA 1600 \AA images with 48 sec cadence over $\pm$ 30 minutes from flare peak and excluded saturated images.
The UV regions are defined as those with the UV emissions over 10$\sigma$ levels.
Figure \ref{fig:ap0} is the averaged light curves for each integration region.
This indicates that the light curves hardly depend on how to define the integration regions and the obtained energies change within a factor of 2, which cannot explain the discrepancy on the $E$-$\tau$ diagram.

As long as our present knowledge and instruments are concerned, we can conclude that the estimated energies and durations of solar WLFs would be realistic and the discrepancy is caused by some kinds of physical reasons.
More investigations are, however, necessary, especially spectroscopic observations of WLFs with broad wavelength coverage.

\clearpage
\section{Validation of the Scaling Law}\label{sec:43}
In Section \ref{dis:cause2}, we proposed the possibility that the discrepancy of $E$-$\tau$ can be explained by the coronal magnetic field strength, and derived the useful equations.
To confirm our suggestions and apply the scaling law to stellar flare observations, we tested the validity of the scaling law on spatially resolved solar flares in our catalogue. 
The coronal magnetic field $B_{\rm obs}$ and loop length $L_{\rm obs}$ are observationally estimated by using images taken with \textit{SDO}/HMI magnetogram and AIA 94 {\AA}, respectively, and we then compared them with theoretical values ($B_{\rm theor}$, $L_{\rm theor}$) obtained from Equations \ref{eq:nam} and \ref{eq:nam2}. 
We simply explain the observational method as below.
Firstly, flaring regions were defined as the regions with the brightness of AIA 94 {\AA} above 50 DN$\rm s^{-1}$. The loop length scales $L_{\rm obs}$ were defined as square roots of the areas of flaring regions.
Secondly, the means of absolute values of photosphetic magnetic field $\bar{B}$ were measured inside the projection of the above flaring area to the photosphere.
Using the empirical relation between coronal and photospheric magnetic field \citep[$B_{\rm corona}\sim B_{\rm photosphere}/3$;][]{2005ApJ...632.1184I,1978SoPh...57..279D}, coronal magnetic field strength $B_{\rm obs}$ were calculated as $\bar{B}/3$. 
Please refer to \cite{2017PASJ...69....7N} for more detail.

\begin{figure}[p]
\begin{center}
\includegraphics[scale=0.5]{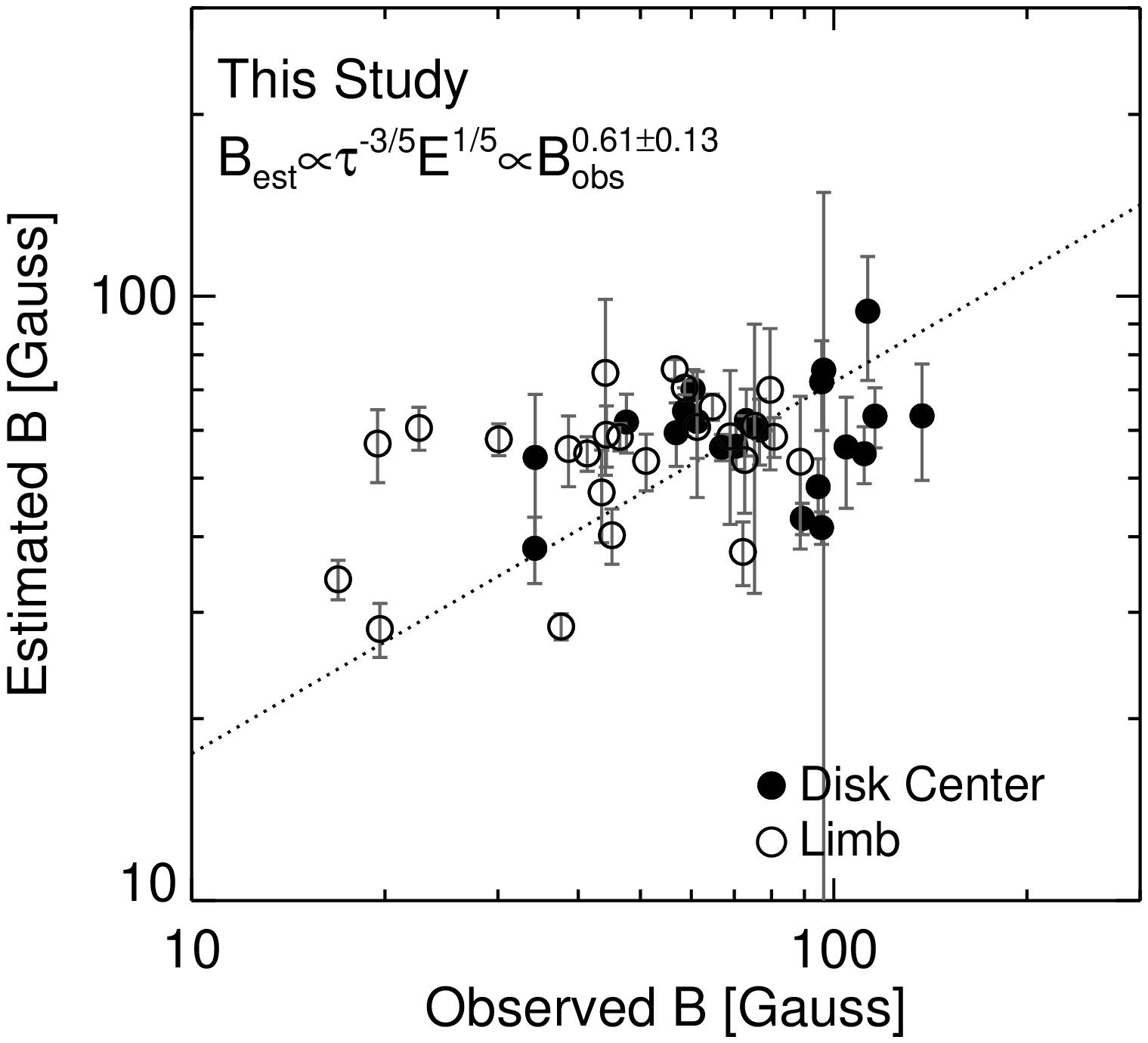}
\includegraphics[scale=0.5]{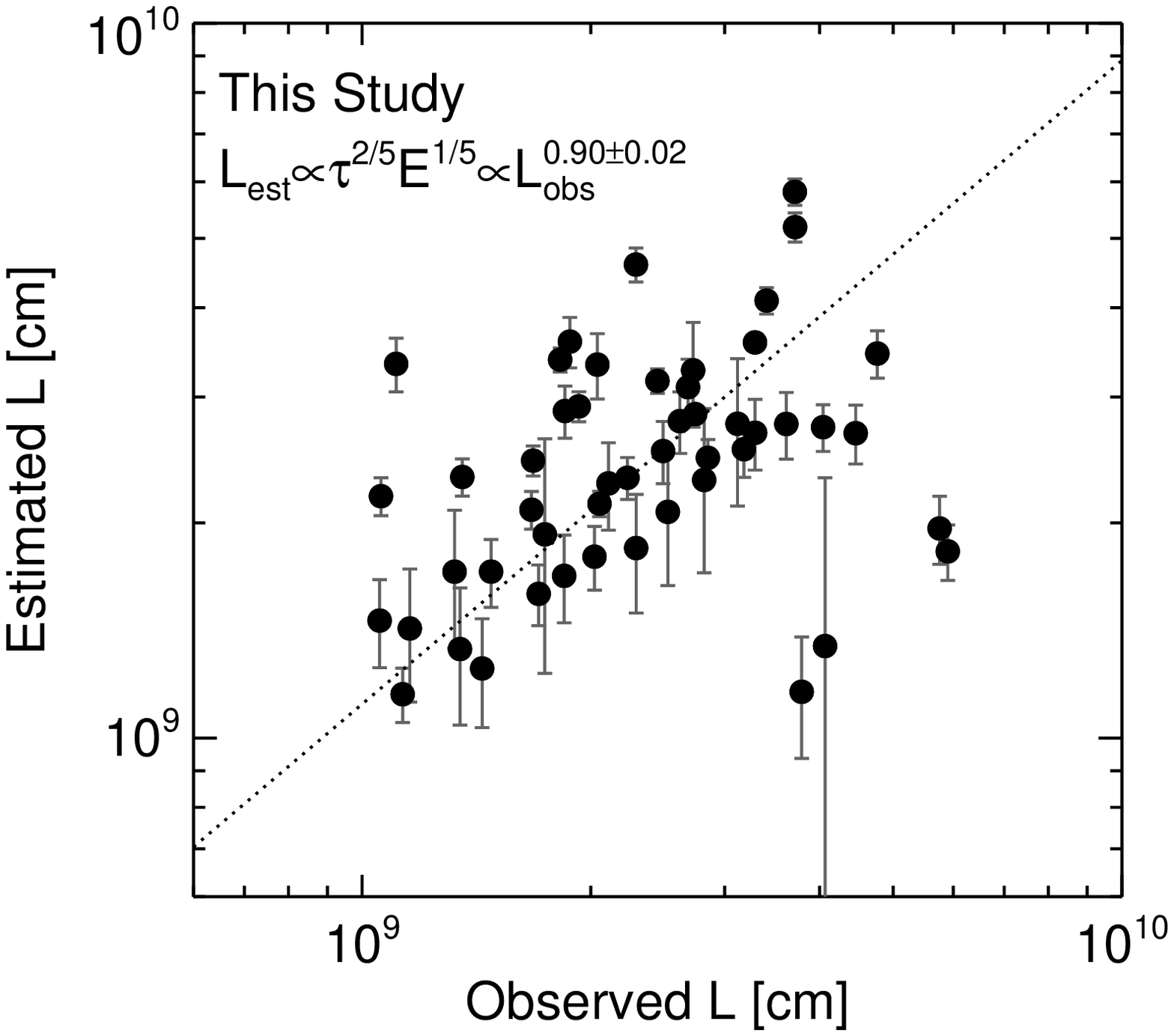}
\caption{Comparisons of the theoretical and observational values. The left and right panels show the comparison of coronal magnetic field strengths $B$ and loop lengths $L$, respectively. In the left panel, open circles are flares on the limb ($>$700) whose magnetic field strength is not reliable. Dotted lines are the fitted lines with linear regression bisecter method. As for the left panel, we fitted the data except for the limb data.}
\label{fig:10}
\end{center}
\end{figure}

The left panel in Figure \ref{fig:10} is a comparison of the theoretical and observational coronal magnetic field strength. 
As we expected, there are weak positive correlations between the theoretical and observational estimated  magnetic fields. 
The failure in the power law index may be caused by the rough method of extrapolation of $B_{\rm obs}$ \citep{2017PASJ...69....7N} and the large scattering would be due to the difficulty in accurate measurements of solar WLFs and the contribution of the filling factor to the scaling laws. 
The right panel in Figure \ref{fig:10} is the same as the left one for the magnetic loop length ($L$).
As one can see, the scaling law can predict the loop length with a linear relation, which is natural because flare energy and durations are basically determined by the length scales in the case of solar flares.

Although there are positive correlations between the observed values and those estimated by our scaling law, 
the validity of the scaling laws cannot be completely confirmed, especially for the magnetic field strength.
This can be because of the difficulty of the measurement of both physical quantities of solar WLFs and coronal magnetic field strength.
As a future study, it is necessary to directly measure stellar magnetic field strengths and compare those estimated from the flare scaling relations.

\clearpage
\section{Applications to Other Studies and Future Stellar Observations}\label{app4}

\begin{figure}[p]
\begin{center}
\includegraphics[scale=0.5]{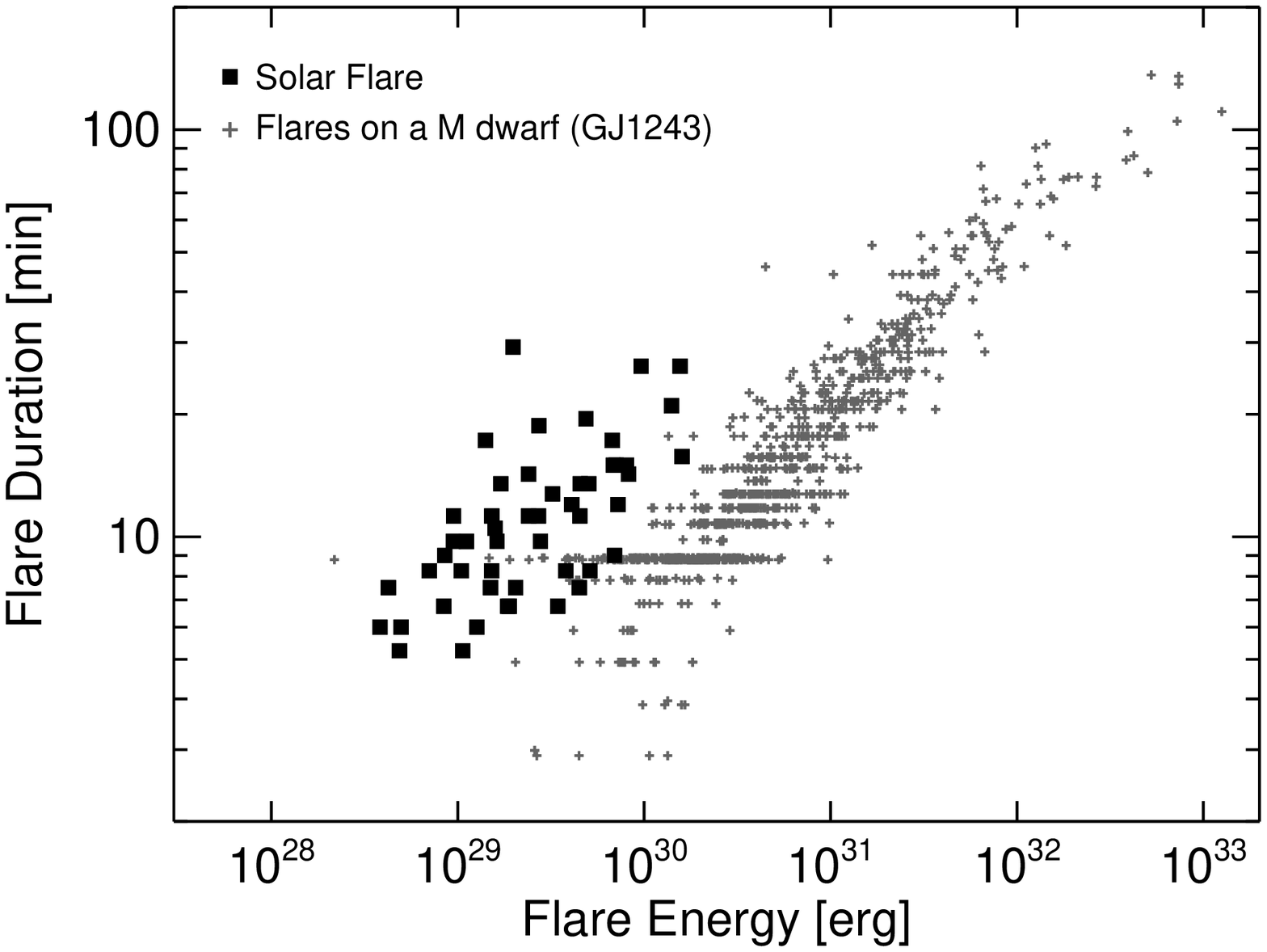}
\caption{A comparison between the flare energies and durations. The filled squares are solar WLFs and the gray crosses are flares on a M-type star GJ1243 \citep{2014ApJ...797..121H}. Note that the durations are defined as that from the beginning to the end of flares, and the energies are converted into that in the $Kepler$ pass band by assuming 10,000 K blackbody}
\label{fig:ap4}
\end{center}
\end{figure}

Here we try to apply our scaling relation (Equation \ref{eq:nam}) to other studies and show our prospects for the future stellar observations.
Firstly we compared the solar WLFs and flares on a M-type star GJ1243 \citep{2014ApJ...797..121H}.
Figure \ref{fig:ap4} is the comparison on the $E$-$\tau$ diagram.
Note that the durations are defined as those from the beginning to the end of flares, and the energies are converted into that in the $Kepler$ pass band by assuming 10,000 K blackbody.
On the basis of our scaling relation (Equation \ref{eq:nam}), the magnetic field strength of GJ 1243 is about 1.7 times stronger than that of solar flares.
This is consistent with our understanding that the magnetic field of M-type stars are stronger than that of the Sun.
However, the scaling law would not be easily applied in this case because the coronal density is expected to be different from solar atmosphere.
Moreover, the emission profile would be expected to be different between solar WLFs and those on M-type stars.
It is therefore necessary to investigate the spectral profiles of solar and stellar WLFs as well as stellar atmospheric parameters in detail.

Next let us apply our scaling laws to the previous papers studying the relation between flare energies (or fluxes) and durations.
For example, \cite{2016PASJ...68...90T} examined the X-ray flare luminosity ($L$) and its duration ($\tau$) about flares on M-type stars, RS CVn stars and the Sun, and discussed the obtained relations $\tau \propto L^{~0.2}$ by using the scaling relation of radiative and conductive cooling.
The theories based of radiative/conductive cooling include an uncertainty because X-ray light curves are observed as superpositions of flaring loops which reconnected one after another.
If the scaling law (Equation \ref{eq:rtv}; $a=4/7$) is applied, the estimated magnetic field strength of flares on M-type stars, RS CVn stars and the Sun is ranging from 50 G to 500 G and predicts the stronger magnetic field strength on M-type and RS CVn stars than on our Sun. 
This is consistent with the observation that M-type and RS CVn stars tend to show extremely high activities \citep[e.g., ][]{2005stam.book.....G} and M-type stars have about three times stronger magnetic field of star spots than our Sun \citep{1996ApJ...459L..95J}.
\cite{2015ApJ...814...35C} moreover carried out a statistical study about WLF energy and duration on mid-M stars. 
The result shows the different distribution on $E$--$\tau$ between nearby M stars (relatively long duration) and open cluster M stars (relatively short duration). 
This is consistent with our understandings that stars in open clusters are young and then have strong magnetic field strength.

As $Kepler$ mission have discovered a huge amount of stellar flares,
the number of such kind of observations would increase in future (e.g., $TESS$; $PLATO$, \citealt{2014ExA....38..249R}). 
In this situation, our scaling law would be helpful for research on stellar activities. 
The advantage of the scaling law is that only by optical photometry we can estimate physical quantities of unresolved stellar surfaces. 

\listofchanges

\begin{thebibliography}{}
\bibitem[Bastien et al.(2016)]{2016ApJ...818...43B} Bastien, F.~A., Stassun, K.~G., Basri, G., \& Pepper, J.\ 2016, \apj, 818, 43 

\bibitem[Carrington(1859)]{1859MNRAS..20...13C} Carrington, R.~C.\ 1859, \mnras, 20, 13 

\bibitem[Chang et al.(2015)]{2015ApJ...814...35C} Chang, S.-W., Byun, Y.-I., \& Hartman, J.~D.\ 2015, \apj, 814, 35 

\bibitem[Christe et al.(2008)]{2008ApJ...677.1385C} Christe, S., Hannah, I.~G., Krucker, S., McTiernan, J., \& Lin, R.~P.\ 2008, \apj, 677, 1385-1394 

\bibitem[Ding et al.(1999)]{1999ApJ...512..454D} Ding, M.~D., Fang, C., \& Yun, H.~S.\ 1999, \apj, 512, 454

\bibitem[Dulk \& McLean(1978)]{1978SoPh...57..279D} Dulk, G.~A., \& McLean, D.~J.\ 1978, \solphys, 57, 279 

\bibitem[Dulk(1985)]{1985ARA&A..23..169D} Dulk, G.~A.\ 1985, \araa, 23, 169 

\bibitem[Emslie et al.(2012)]{2012ApJ...759...71E} Emslie, A.~G., Dennis, B.~R., Shih, A.~Y., et al.\ 2012, \apj, 759, 71 

\bibitem[Fletcher \& Hudson(2008)]{2008ApJ...675.1645F} Fletcher, L., \& Hudson, H.~S.\ 2008, \apj, 675, 1645-1655 

\bibitem[Gershberg(2005)]{2005stam.book.....G} Gershberg, R.~E.\ 2005, Studies in Applied Mathematics,  

\bibitem[Grosso et al.(1997)]{1997Natur.387...56G} Grosso, N., Montmerle, T., Feigelson, E.~D., et al.\ 1997, \nat, 387, 56 

\bibitem[Hawley \& Fisher(1992)]{1992ApJS...78..565H} Hawley, S.~L., \& Fisher, G.~H.\ 1992, \apjs, 78, 565 

\bibitem[Hawley et al.(2014)]{2014ApJ...797..121H} Hawley, S.~L., Davenport, J.~R.~A., Kowalski, A.~F., et al.\ 2014, \apj, 797, 121 

\bibitem[Heinzel \& Kleint(2014)]{2014ApJ...794L..23H} Heinzel, P., \& Kleint, L.\ 2014, \apjl, 794, L23

\bibitem[Heinzel et al.(2017)]{2017ApJ...847...48H} Heinzel, P., Kleint, L., Ka{\v s}parov{\'a}, J., \& Krucker, S.\ 2017, \apj, 847, 48

\bibitem[Hudson et al.(1992)]{1992PASJ...44L..77H} Hudson, H.~S., Acton, L.~W., Hirayama, T., \& Uchida, Y.\ 1992, \pasj, 44, L77 

\bibitem[Hudson et al.(2006)]{2006SoPh..234...79H} Hudson, H.~S., Wolfson, C.~J., \& Metcalf, T.~R.\ 2006, \solphys, 234, 79 

\bibitem[Hurford et al.(2002)]{2002SoPh..210...61H} Hurford, G.~J., Schmahl, E.~J., Schwartz, R.~A., et al.\ 2002, \solphys, 210, 61 

\bibitem[Ishii et al.(2013)]{2013PASJ...65...39I} Ishii, T.~T., Kawate, T., Nakatani, Y., et al.\ 2013, \pasj, 65, 39 

\bibitem[Isobe et al.(1990)]{1990ApJ...364..104I} Isobe, T., Feigelson, E.~D., Akritas, M.~G., \& Babu, G.~J.\ 1990, \apj, 364, 104 

\bibitem[Isobe et al.(2005)]{2005ApJ...632.1184I} Isobe, H., Takasaki, H., \& Shibata, K.\ 2005, \apj, 632, 1184 

\bibitem[Isobe et al.(2007)]{2007PASJ...59S.807I} Isobe, H., Kubo, M., Minoshima, T., et al.\ 2007, \pasj, 59, S807 

\bibitem[Jess et al.(2008)]{2008ApJ...688L.119J} Jess, D.~B., Mathioudakis, M., Crockett, P.~J., \& Keenan, F.~P.\ 2008, \apjl, 688, L119 

\bibitem[Johns-Krull \& Valenti(1996)]{1996ApJ...459L..95J} Johns-Krull, C.~M., \& Valenti, J.~A.\ 1996, \apjl, 459, L95 

\bibitem[Katsova \& Livshits(2015)]{2015SoPh..290.3663K} Katsova, M.~M., \& Livshits, M.~A.\ 2015, \solphys, 290, 3663

\bibitem[Kawate et al.(2016)]{2016ApJ...833...50K} Kawate, T., Ishii, T.~T., Nakatani, Y., et al.\ 2016, \apj, 833, 50 

\bibitem[Kerr \& Fletcher(2014)]{2014ApJ...783...98K} Kerr, G.~S., \& Fletcher, L.\ 2014, \apj, 783, 98 

\bibitem[Kleint et al.(2016)]{2016ApJ...816...88K} Kleint, L., Heinzel, P., Judge, P., \& Krucker, S.\ 2016, \apj, 816, 88 

\bibitem[Kowalski et al.(2017)]{2017ApJ...837..125K} Kowalski, A.~F., Allred, J.~C., Uitenbroek, H., et al.\ 2017, \apj, 837, 125 

\bibitem[Kuhar et al.(2016)]{2016ApJ...816....6K} Kuhar, M., Krucker, S., Mart{\'{\i}}nez Oliveros, J.~C., et al.\ 2016, \apj, 816, 6 

\bibitem[Kretzschmar(2011)]{2011A&A...530A..84K} Kretzschmar, M.\ 2011, \aap, 530, A84 

\bibitem[Krucker et al.(2011)]{2011ApJ...739...96K} Krucker, S., Hudson, H.~S., Jeffrey, N.~L.~S., et al.\ 2011, \apj, 739, 96 

\bibitem[Lemen et al.(2012)]{2012SoPh..275...17L} Lemen, J.~R., Title, A.~M., Akin, D.~J., et al.\ 2012, \solphys, 275, 17 

\bibitem[Lin \& Hudson(1976)]{1976SoPh...50..153L} Lin, R.~P., \& Hudson, H.~S.\ 1976, \solphys, 50, 153 

\bibitem[Lin et al.(2002)]{2002SoPh..210....3L} Lin, R.~P., Dennis, B.~R., Hurford, G.~J., et al.\ 2002, \solphys, 210, 3 

\bibitem[Machado et al.(1986)]{1986A&A...159...33M} Machado, M.~E., Emslie, A.~G., \& Mauas, P.~J.\ 1986, \aap, 159, 33 

\bibitem[Machado et al.(1989)]{1989SoPh..124..303M} Machado, M.~E., Emslie, A.~G., \& Avrett, E.~H.\ 1989, \solphys, 124, 303 

\bibitem[Maehara et al.(2012)]{2012Natur.485..478M} Maehara, H., Shibayama, T., Notsu, S., et al.\ 2012, \nat, 485, 478 

\bibitem[Maehara et al.(2015)]{2015EP&S...67...59M} Maehara, H., Shibayama, T., Notsu, Y., et al.\ 2015, Earth, Planets, and Space, 67, 59 

\bibitem[Maehara et al.(2017)]{2017PASJ...69...41M} Maehara, H., Notsu, Y., Notsu, S., et al.\ 2017, \pasj, 69, 41 

\bibitem[Mart{\'{\i}}nez Oliveros et al.(2012)]{2012ApJ...753L..26M} Mart{\'{\i}}nez Oliveros, J.-C., Hudson, H.~S., Hurford, G.~J., et al.\ 2012, \apjl, 753, L26 

\bibitem[Matthews et al.(2003)]{2003A&A...409.1107M} Matthews, S.~A., van Driel-Gesztelyi, L., Hudson, H.~S., \& Nitta, N.~V.\ 2003, \aap, 409, 1107 

\bibitem[Mathur et al.(2017)]{2017ApJS..229...30M} Mathur, S., Huber, D., Batalha, N.~M., et al.\ 2017, \apjs, 229, 30 

\bibitem[Mauas(1990)]{1990ApJS...74..609M} Mauas, P.~J.~D.\ 1990, \apjs, 74, 609

\bibitem[Najita \& Orrall(1970)]{1970SoPh...15..176N} Najita, K., \& Orrall, F.~Q.\ 1970, \solphys, 15, 176 

\bibitem[Namekata et al.(2017)]{2017PASJ...69....7N} Namekata, K., Sakaue, T., Watanabe, K., Asai, A., \& Shibata, K.\ 2017, \pasj, 69, 7 

\bibitem[Neidig(1989)]{1989SoPh..121..261N} Neidig, D.~F.\ 1989, \solphys, 121, 261 

\bibitem[Neupert(1968)]{1968ApJ...153L..59N} Neupert, W.~M.\ 1968, \apjl, 153, L59 

\bibitem[Notsu et al.(2013)]{2013ApJ...771..127N} Notsu, Y., Shibayama, T., Maehara, H., et al.\ 2013, \apj, 771, 127 

\bibitem[Notsu et al.(2015)]{2015PASJ...67...33N} Notsu, Y., Honda, S., Maehara, H., et al.\ 2015, \pasj, 67, 33

\bibitem[Priest(1981)]{1981sfmh.book.....P} Priest, E.~R.\ 1981, Solar Flare Magnetohydrodynamics  New York; Gordon and Breach Science Publishers)

\bibitem[Ricker et al.(2015)]{2015JATIS...1a4003R} Ricker, G.~R., Winn, J.~N., Vanderspek, R., et al.\ 2015, Journal of Astronomical Telescopes, Instruments, and Systems, 1, 014003 

\bibitem[Rosner et al.(1978)]{1978ApJ...220..643R} Rosner, R., Tucker, W.~H., \& Vaiana, G.~S.\ 1978, \apj, 220, 643 

\bibitem[Rauer et al.(2014)]{2014ExA....38..249R} Rauer, H., Catala, C., Aerts, C., et al.\ 2014, Experimental Astronomy, 38, 249 

\bibitem[Rust \& Bar(1973)]{1973SoPh...33..445R} Rust, D.~M., \& Bar, V.\ 1973, \solphys, 33, 445

\bibitem[Scherrer et al.(2012)]{2012SoPh..275..207S} Scherrer, P.~H., Schou, J., Bush, R.~I., et al.\ 2012, \solphys, 275, 207 

\bibitem[Shibata \& Yokoyama(1999)]{1999ApJ...526L..49S} Shibata, K., \& Yokoyama, T.\ 1999, \apjl, 526, L49 

\bibitem[Shibata \& Yokoyama(2002)]{2002ApJ...577..422S} Shibata, K., \& Yokoyama, T.\ 2002, \apj, 577, 422 

\bibitem[Shibata \& Magara(2011)]{2011LRSP....8....6S} Shibata, K., \& Magara, T.\ 2011, Living Reviews in Solar Physics, 8, 6 

\bibitem[Shibayama et al.(2013)]{2013ApJS..209....5S} Shibayama, T., Maehara, H., Notsu, S., et al.\ 2013, \apjs, 209, 5 

\bibitem[Svestka(1976)]{1976sofl.book.....S} Svestka, Z.\ 1976, Solar Flares.~Svestka, Z., pp.~415.~ISBN 90-277-0662-X.~Springer-Verlag Berlin Heidelberg 1976, 415

\bibitem[Takahashi et al.(2016)]{2016ApJ...833L...8T} Takahashi, T., Mizuno, Y., \& Shibata, K.\ 2016, \apjl, 833, L8 

\bibitem[Toriumi et al.(2017)]{2017ApJ...834...56T} Toriumi, S., Schrijver, C.~J., Harra, L.~K., Hudson, H., \& Nagashima, K.\ 2017, \apj, 834, 56

\bibitem[Tsuboi et al.(2016)]{2016PASJ...68...90T} Tsuboi, Y., Yamazaki, K., Sugawara, Y., et al.\ 2016, \pasj, 68, 90 

\bibitem[Tsuneta(1996)]{1996ApJ...456..840T} Tsuneta, S.\ 1996, \apj, 456, 840

\bibitem[Tsuneta et al.(2008)]{2008SoPh..249..167T} Tsuneta, S., Ichimoto, K., Katsukawa, Y., et al.\ 2008, \solphys, 249, 167 

\bibitem[Veronig et al.(2002)]{2002A&A...382.1070V} Veronig, A., Temmer, M., Hanslmeier, A., Otruba, W., \& Messerotti, M.\ 2002, \aap, 382, 1070 

\bibitem[Watanabe et al.(2013)]{2013ApJ...776..123W} Watanabe, K., Shimizu, T., Masuda, S., Ichimoto, K., \& Ohno, M.\ 2013, \apj, 776, 123 

\bibitem[Watanabe et al.(2017)]{2017arXiv171009531W} Watanabe, K., Kitagawa, J., \& Masuda, S.\ 2017, arXiv:1710.09531

\bibitem[Wright et al.(2011)]{2011ApJ...743...48W} Wright, N.~J., Drake, J.~J., Mamajek, E.~E., \& Henry, G.~W.\ 2011, \apj, 743, 48 

\bibitem[Xu et al.(2006)]{2006ApJ...641.1210X} Xu, Y., Cao, W., Liu, C., et al.\ 2006, \apj, 641, 1210 


\end{thebibliography}
\end{document}